\documentclass[12pt]{iopart}
\usepackage{cite}
\usepackage{amsmath,amssymb,amsfonts}
\usepackage{algorithm}
\usepackage{graphicx}
\usepackage{textcomp}
\usepackage{multirow}
\usepackage{xcolor}
\usepackage{doi}
\usepackage{comment}
\usepackage{braket}
\usepackage{booktabs}
\usepackage{pifont}
\usepackage{todonotes}
\usepackage{siunitx}
\usepackage{bm}
\usepackage{subcaption}
\usepackage{algpseudocode}
\usepackage{fixme}
\usepackage{amsthm}
\usepackage{textcomp}

\newcommand{\ev}[2]{\langle #2|#1|#2\rangle}
\newcommand{\dyad}[1]{ |#1\rangle\langle#1|}

\fxsetup{status=draft} 

 \newtheorem{definition}{Definition}

\begin{document}
\title{GALIC: Hybrid Multi-Qubitwise Pauli Grouping for Quantum Computing Measurement} 
\thispagestyle{plain}
\pagestyle{plain}
\author{Matthew X. Burns$^1$$^2$, Chenxu Liu$^1$,
Samuel Stein$^1$, Bo Peng$^1$, Karol Kowalski$^1$, and Ang Li$^{13}$}
\address{$^1$ Physical and Computational Sciences, Pacific Northwest National Laboratory, Richland, WA, 99354, USA}
\address{$^2$ Department of Electrical and Computer Engineering, University of Rochester. 500 Computer Studies Building
P.O. Box 270231
Rochester, NY 14627, USA}

\address{$^3$ Department of Electrical and Computer Engineering, University of Washington, 1410 NE Campus Pkwy, Seattle, WA 98195, USA}

\eads{\mailto{mburns13@ur.rochester.edu}, \mailto{ang.li@pnnl.gov}}

\begin{abstract}
\noindent Observable estimation is a core primitive in NISQ-era algorithms targeting quantum chemistry applications. To reduce the state preparation overhead required for accurate estimation, recent works have proposed various simultaneous measurement schemes to lower estimator variance. Two primary grouping schemes have been proposed: fully commutativity (FC) and qubit-wise commutativity (QWC), with no compelling means of interpolation. In this work we propose a generalized framework for designing and analyzing context-aware hybrid FC/QWC commutativity relations. We use our framework to propose a noise-and-connectivity aware grouping strategy: Generalized backend-Aware pauLI Commutation (GALIC). We demonstrate how GALIC interpolates between FC and QWC, maintaining estimator accuracy in Hamiltonian estimation while lowering variance by an average of 20\% compared to QWC. We also explore the design space of near-term quantum devices using the GALIC framework, specifically comparing device noise levels and connectivity. We find that error suppression has a more than $10\times$ larger impact on device-aware estimator variance than qubit connectivity with even larger correlation differences in estimator biases.

\end{abstract}
\section{Introduction}\label{sec:introduction}

Quantum Hamiltonian simulation  promises order-of-magnitude speedups over classical methods in both near and far term systems. Fault tolerant systems achieve exponential gains in quantum simulation~\cite{babbush_low-depth_2018,simon_improved_2024} via Quantum Phase Estimation (QPE)~\cite{nielsen_quantum_2011} and qubitization~\cite{low_hamiltonian_2019}. NISQ-era systems offer potential advantage through shallow-circuit algorithms such as the Variational Quantum Eigensolver (VQE)~\cite{peruzzo_variational_2014,tilly_variational_2022} which leverage quantum entanglement and vast Hilbert space for expectation value estimation beyond the reach of classical resources.

Although \textit{feasible} on near-term devices, NISQ algorithm efficiency is fundamentally limited by the measurement overhead of expectation value estimation~\cite{gonthier_measurements_2022}. The number of distinct $N$-qubit operator expectations to estimate scales $O(N^4)$ for VQE, with adaptive algorithms such as ADAPT-VQE~\cite{grimsley_adaptive_2019} scaling $O(N^6)$~\cite{liu_efficient_2021} in optimized implementations and up to $O(N^8)$. Each operator requires thousands of measurements~\cite{zhu_optimizing_2024, gonthier_measurements_2022, yen_deterministic_2023}, requiring millions of state preparations to obtain Hamiltonian energy estimates within chemical accuracy~\cite{yen_deterministic_2023}.

Among the most popular measurement reduction methods are ``simultaneous measurement'' schemes, in which a single measurement contributes to multiple expectation value estimates.
Two primary simultaneous measurement schemes have been proposed in literature. Classical Shadow tomography~\cite{huang_predicting_2020,hu_classical_2021,koh_classical_2022,hillmich_decision_2021,elben_randomized_2022} and related methods~\cite{wu_overlapped_2023,elben_randomized_2022} construct expectation value estimates using random basis sampling. An alternative approach is Pauli grouping~\cite{yen_deterministic_2023,crawford_efficient_2021,gokhale_optimization_2020,gokhale2020n,choi_improving_2022,bansingh_fidelity_2022,miller2022hardware,escudero_hardware-efficient_2023}, which simultaneously measures sets of commuting operators from the target observable, enabling the construction of lower-variance estimators. Initial works focused largely on combinatorial optimization algorithms for constructing commuting groups, with a particular emphasis on solving NP-Hard Minimum Clique Cover reductions using heuristic methods~\cite{verteletskyi_measurement_2020,gokhale_optimization_2020,gokhale2020n,crawford_efficient_2021}. 
Efforts have lately focused on advanced grouping algorithms with even lower measurement overhead: including iterative optimization by adaptive variance estimation~\cite{wu_overlapped_2023,yen_deterministic_2023,shlosberg_adaptive_2023}, overlapping commuting groups~\cite{wu_overlapped_2023,yen_deterministic_2023}, and canceling variance terms with additional commuting operators~\cite{choi_improving_2022}. Recent works have shown that these adaptive grouping strategies offer lower measurement overhead than tomography~\cite{yen_deterministic_2023,choi_improving_2022} motivating their use for near-term quantum advantage. Two main commutation relations have been used for grouping proposals, fully commuting (FC)~\cite{gokhale_optimization_2020,gokhale2020n,crawford_efficient_2021,yen_deterministic_2023,choi_improving_2022} and qubit-wise commuting (QWC)~\cite{qiskit2024,mcclean_openfermion_2019,romero_strategies_2018,yen_deterministic_2023}.

Existing Pauli grouping works treat the choice between FC and QWC as a rigid binary. While FC Pauli groups offer the lowest estimator variances~\cite{yen_deterministic_2023,gokhale_optimization_2020,crawford_efficient_2021}, they require heavily entangled diagonalization circuits~\cite{crawford_efficient_2021}. Noisy FC entangling operations contribute significant estimator bias~\cite{escudero_hardware-efficient_2023, miller2022hardware}, requiring additional measurement cost of quantum error mitigation~\cite{cai_quantum_2023}. QWC was proposed as a near-term alternative, requiring no entangling operations~\cite{gokhale2020n,romero_strategies_2018}. The ease of implementation and low-depth measurement circuits have led to wide adoption by quantum software packages~\cite{qiskit2024,mcclean_openfermion_2019,kottmann_tequila_2021}. QWC pays for high fidelity with significantly increased variance (and thereby measurement cost) compared to FC~\cite{yen_deterministic_2023,gokhale_optimization_2020,choi_improving_2022}. In an exception to the overall trend, some recent works have offered a hardware-efficient grouping strategy which permits \textit{some} entangling gates~\cite{miller2022hardware,escudero_hardware-efficient_2023}. However, these are distinct and isolated efforts, with no overarching framework for designing hybrid QWC/FC strategies.

In this work, we provide two major contributions:
\begin{enumerate}
    \item We propose a general framework of QWC/FC interpolation, opening a previously unexplored design space for measurement optimization. 
    \item Using our framework, we design a novel measurement strategy which considers both device fidelity and connectivity: the Generalized backend-Aware pauLI Commutativity (GALIC) scheme
\end{enumerate}
We evaluate GALIC against existing grouping strategies with extensive numerical simulations on 5 molecular Hamiltonians up to 14 qubits using device models from IBM and IonQ. GALIC estimators are shown to maintain chemical accuracy ($<1$ kcal/mol error) while achieving $>20\%$ shot overhead reduction over QWC. Experiments on an IBM quantum processor further show 1.2$\times$ lower GALIC estimator variance compared to QWC, validating our numerical results. Using GALIC, we numerically explore the impact of NISQ device connectivity and fidelity on shot overhead and estimator bias. We find that the correlation between device fidelity and estimator variance is 10.6$\times$ greater than connectivity, with similar results for energy estimation bias.  We therefore conclude that gate fidelity improvement is a more promising pathway for near-term quantum advantage compared with increased qubit connectivity.
\section{Background}\label{sec:background}
In this section we first formalize the task of quantum observable estimation, then motivate grouping strategies for simultaneous measurement. Fully commuting (FC) and qubit-wise commuting (QWC) groups are compared using common statistical figures of merit. We discuss the shortcomings of each method, motivating the development of a hybrid approach.
\renewcommand{\tr}[1]{\mathrm{Tr}[#1]}
\newcommand{\E}[1]{\mathbb{E}[#1]}
\newcommand{\Var}[1]{\mathrm{Var}[#1]}
\subsection{Quantum Measurement}
Suppose we have an observable of interest $H$ which can be decomposed into individual Pauli observables $P_i$:
\begin{equation}\label{eqn:pauli_hamil}
    H=\sum_{i=1}^Mc_iP_i,
\end{equation}
where each $P_i=\bigotimes_{j=1}^N \sigma_{ij}$ is an $N$-qubit tensor product over $M$ individual Pauli operators $\sigma_{ij}\in\{X,Y,Z,I\}$ with coefficients $c_i\in\mathbb{R}$. For classically intractable systems, we cannot compute the expectation value $\ev{H}{\psi}$ analytically, requiring quantum sampling. To measure $\ev{H}{\psi}$ on a quantum computer, we construct an \emph{estimator} $\hat{H}$ which seeks to approximate $\ev{H}{\psi}$ by estimating each $\ev{P_i}{\psi}$ and computing Eqn.~\eqref{eqn:pauli_hamil}.

To generalize our discussion to noisy devices, we switch from this point to a density matrix formalism. An ideal, pure quantum state $\psi$ is expressed in density matrix form as $\rho=\dyad{\psi}$. We then have $\ev{H}{\psi}=\tr{H\rho}$. From Eqn.~\eqref{eqn:pauli_hamil}, we construct estimator $\hat{H}$ as:
\begin{equation}
    \hat{H}=\sum_{i=1}c_i\hat{P_i},
\end{equation}
where each estimator $\hat{P_i}$ approximates $\tr{P_i\rho}$. Here, we assume $\hat{P_i}$ to be the sample mean estimator, given by: 
\begin{equation}
    \hat{P}_i=\frac{1}{n_i}\sum_{m=1}^{n_i}p_i^m
\end{equation}
where $n_i$ is the number of $P_i$ samples taken and each $p_i^m\in\{-1,1\}$ is a single measurement result. Measuring an arbitrary Pauli product $P_{i}$ requires a unitary transformation $U_{i}$ to rotate $P_{i}$ to the computational basis: a ``measurement circuit''. Applying a unitary $U_i$ leaves the expectation value invariant, as $\tr{P_iU_{i}\rho U_{i}^\dagger}=\tr{U_{i}^\dagger P_iU_{i}\rho }$. The resulting operator $U_{i}^\dagger P_iU_{i}=\bigotimes Z^{(j)}$ lies within the computational basis, making measurements possible.
A single Pauli string can be rotated to the computational basis using single-qubit Clifford operations~\cite{nielsen_quantum_2011,gokhale_optimization_2020}, making single-Pauli $U_i$ trivially implementable.

\subsection{Operator Grouping}
In the naive case, we construct a measurement circuit for each $P_i$, perform the rotation, and perform a single measurement $p_i^m$. That is, we obtain one measurement for each shot. In applications targeting near-term systems, such as the variational quantum eigensolver (VQE)~\cite{peruzzo_variational_2014,tilly_variational_2022}, the number of Pauli terms ($M$) scales $O(N^4)$, where $N$ is the number of spin orbitals (a.k.a. qubits) in the problem of interest. In the case of adaptive schemes with commutator-based gradients, the number of individual Pauli operators scales $O(N^6)$~\cite{liu_efficient_2021} and up to $O(N^8)$~\cite{grimsley_adaptive_2019}. 

To lower the shot complexity of observable estimation, several works have proposed Pauli grouping schemes~\cite{gokhale2020n,gokhale_optimization_2020,yen_deterministic_2023,miller2022hardware,escudero_hardware-efficient_2023} which allow for the simultaneous measurement of several Pauli observables and reducing estimation complexity by up to a factor of $N$~\cite{gokhale2020n}. Simultaneous measurement can be applied to any Pauli-decomposed operator, and is therefore compatible with other measurement reduction strategies such as Hamiltonian modification~\cite{zhao_measurement_2020, ralli_unitary_2023} and engineered likelihood functions~\cite{wang_minimizing_2021}.

Simultaneous measurement permits a single state preparation to provide measurements for several distinct $\{P_i\}$. To enable simultaneous measurement, we require the operators $\{P_i\}$ to pairwise \emph{commute}, i.e., $[P_i,P_j]=P_iP_j-P_jP_i=0$. If two operators commute, they can be diagonalized in the same basis~\cite{nielsen_quantum_2011}, meaning that we can measure both operators simultaneously. 

\begin{figure}
    \centering
    \includegraphics[width=\linewidth]{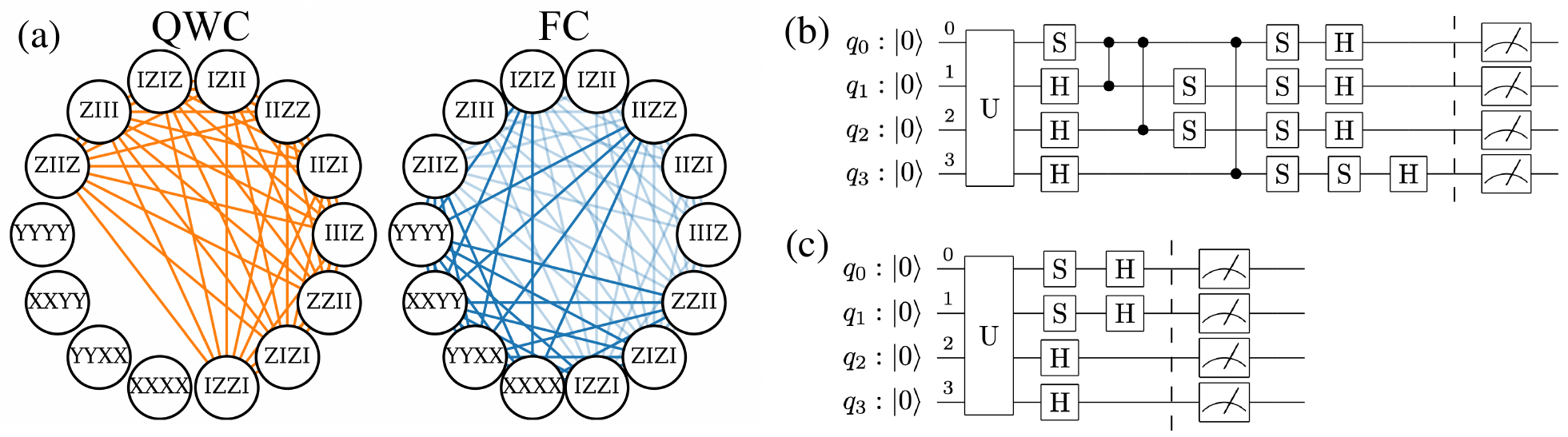}
    \caption{(a) Illustration of commutation relations for Pauli strings from the Jordan-Wigner~\cite{wigner1928paulische} mapped $\mathrm{H_2}$ Hamiltonian, where edges denote commutativity. Desaturated edges on the FC graph are also allowed under QWC. (b) The measurement circuit for the FC group $\{$YYYY, XXYY, YYXX, XXXX$\}$, built using the ``CZ'' algorithm described in Ref.~\cite{crawford_efficient_2021}. (c) The measurement circuit for the QWC group $\{$XXII, XIYY$\}$.}
    \label{fig:commute_h2}
\end{figure}

Two primary commutation criteria have been proposed for simultaneous measurement: full commutativity (FC) and qubit-wise commutativity (QWC). FC allows for larger commuting groups at the expense of requiring entangling gates in measurement unitaries $U_i$. QWC is a much more restricted relation, however, it eschews two-qubit gates entirely, making it much more amenable to NISQ-era hardware. Fig.~\ref{fig:commute_h2} illustrates the FC and QWC commutation relations on the Jordan-Wigner~\cite{wigner1928paulische} mapped STO-3G $\mathrm{H_2}$ Hamiltonian~\cite{sun2018pyscf,mcardle_quantum_2020}.

\subsubsection{Fully Commuting Groups}
The most general commutativity relation for Pauli operators is \emph{full commutativity} (FC)~\footnote{Also called ``general commutativity''~\cite{gokhale_optimization_2020}}. As long as two Pauli strings commute, they can be grouped to be measured simultaneously. For example, as $[XXYY, YYYY] = 0$, they can be grouped and measured together in the $H_2$ example. Similarly, ZIII and ZZII can be measured simultaneously. However, as $[ZIII, XXYY] \neq 0$, these two terms cannot be grouped.

Diagonalizing an FC group with anti-commuting qubits requires entangling gates between anti-commuting pairs. Luckily, the transformation diagonalizing commuting Pauli strings in the computational basis is Clifford, which can be efficiently solved using classical stabilizer simulators~\cite{aaronson_improved_2004}(See Ref.~\cite{crawford_efficient_2021} for a pedagogical explanation of circuit construction).
Fig.~\ref{fig:commute_h2}(b) shows the measurement circuit for the FC group $\{\mathrm{YYYY}, \mathrm{XXYY}, \mathrm{YYXX}, \mathrm{XXXX}\}$ which uses three CZ gates.

As FC is the most general commuting scheme, it results in the fewest commuting groups for a given operator pool and the lowest-variance simultaneous measurement estimators~\cite{crawford_efficient_2021,yen_deterministic_2023,gokhale_optimization_2020}. However, FC circuits need $\frac{N_{AC}(N_{AC}-1)}{2}$ entangling gates in the worst case for $N_{AC}$ qubits~\cite{crawford_efficient_2021,van_den_nest_graphical_2004}. Entangling operations are a significant source of error on near-term devices~\cite{gao_practical_2021, bansingh_fidelity_2022}, with error rates an order of magnitude greater than single-qubit gates. Moreover, sparse quantum processing unit (QPU) architectures may not natively support the required unitary entanglements. FC circuits then often incur routing overhead, requiring additional two-qubit gates to perform SWAP operations~\cite{li_tackling_2019}. 

\newcommand{\X}{\mathrm{X}}
\newcommand{\XX}{\mathrm{XX}}
\newcommand{\Y}{\mathrm{Y}}
\newcommand{\YY}{\mathrm{YY}}
\newcommand{\Z}{\mathrm{Z}}
\newcommand{\ZZ}{\mathrm{ZZ}}
\newcommand{\II}{\mathrm{II}}
\newcommand{\I}{\mathrm{I}}
\subsubsection{Qubit-Wise Commuting Groups}
In response to NISQ-era difficulties with FC measurement circuits, qubit-wise commuting (QWC) schemes were proposed to leverage simultaneous measurement with lower circuit overhead. Also called ``tensor product basis'' (TPB) commutation~\cite{escudero_hardware-efficient_2023,mcclean_openfermion_2019,hamamura_efficient_2020}, QWC requires that \textit{no} qubits anti-commute. For example, XXYY and XIIY qubit-wise commute, as all pairwise qubit operators commute. However, XXZZ and ZZXX do not qubit-wise commute, despite $[\XX\ZZ,\ZZ\XX]=0$.

The advantage of QWC lies in its simple measurement unitaries: no entangling operation is needed, i.e., Pauli strings in the same group can be rotated to the computational basis using single-qubit operations. Fig.~\ref{fig:commute_h2}(c) shows QWC for the group $\{\X\X\I\I, \X\I\Y\Y\}$, where X operators are rotated using a Hadamard gate and Y operators are rotated using a Hadamard composed with a phase gate. Note that measuring any QWC group is equivalent to measuring one Pauli string with identity operators superseded by non-identity terms. For example, the measurement circuit for  $\{\X\X\I\I, \X\I\Y\Y\}$ is equivalent to the circuit for XXYY. QWC measurement circuits are therefore no more expensive than naive Pauli measurement, providing NISQ-amenable simultaneous measurement.

\subsection{Grouping Figures of Merit}\label{ssec:metrics}
Evaluating the performance of a grouping scheme requires evaluating the performance of its associated estimator $\hat H$. The most common figure of merit is the mean squared error (MSE):
\begin{equation}
\begin{split}
\label{eqn:mse}
    \mathrm{MSE}(\hat{H}, H)&=\mathbb{E}[(\hat{H}- \tr{H\rho})^2]=\overbrace{(\mathbb{E}[\hat{H}]- \tr{H\rho})^2}^{\textrm{Bias}^2}+\overbrace{\mathbb{E}[\hat{H}^2]- \mathbb{E}[\hat{H}]^2}^{\textrm{Variance}},
\end{split}
\end{equation}
where we have performed the standard bias/variance decomposition. Bias describes the proximity of the estimator mean to the true value (``accuracy'') while the variance describes the magnitude of fluctuations about that mean (``precision'').

In an ideal operation where there are no errors, the expectation value can be estimated with no bias. However, when errors and imperfections in quantum hardware are taken into consideration, the measurement transformations will no longer be perfect, and the density operator is transformed to $\tilde{\rho}_i = \mathcal{E}_{U_i}(\rho)$, where the $\mathcal{E}$ represent the quantum channel with noise and imperfections. For example, if the measurement gate implementation induces a depolarization channel with error probability $p$, the channel is 
\begin{equation}
    \mathcal{E}_{U_i}(\rho) = (1-p) U_i \rho U_i^\dagger + p I/2^N, 
\end{equation}
where $I$ is the identity operator, and $N$ is the number of qubits.

With noise factored in, operator $P_i$ has expectation and variance: $\mathbb{E}[P_i]=\tr{P_i\tilde{\rho}_i}$ and $\Var{P_i}=\tr{P_i^2\tilde{\rho}_i}-\tr{P_i\tilde{\rho}_i}^2$, respectively. The estimator expectation value is fixed $\E{\hat P_i}=\E{P_i}$, while in both ideal and noisy settings the estimator variance decays linearly in the number of samples $\Var{\hat P_i}=\frac{1}{n_i}\Var{P_i}$\footnote{However, $\Var{P_i}$ may differ between the noisy and noiseless cases~\cite{bansingh_fidelity_2022}}. Note that the degree of error in $\tilde \rho_i$ depends on the grouping scheme. Entanglement-heavy FC circuits lead to significantly larger biases than shallow QWC circuits~\cite{escudero_hardware-efficient_2023,bansingh_fidelity_2022}.

For naive single Pauli measurement, we have:
\begin{align}
\textrm{Bias:} \; \mathbb{E}[\hat{H}]- \tr{H\rho} & =\sum_{i=1}^Mc_i(\tr{P_i\tilde{\rho}_i}-\tr{P_i\rho}), \nonumber \\
\textrm{Variance:}\; \Var{\hat{H}} & =\sum_{i=1}^M\frac{|c_i|^2}{n_i}\Var{P_i}, \nonumber
\end{align}
where the additivity of the variance follows from the independence of each estimator $\hat P_i$.

Simultaneous measurement schemes group Pauli operators into $L$ sets $O_\ell=\{P_{\ell j}\}_{j=1}^{M_\ell}$, each containing $M_\ell$ operators which can be simultaneously diagonalized using a single measurement unitary. The noise added by the measurement circuit results in the mixed state $\rho_\ell$. The estimators $\hat{P}_{\ell j},\;\hat P_{\ell i }$ are correlated, however each group estimator $\hat{O_\ell}=\sum_{j=1}^{M_\ell}c_jP_{\ell j}$ is independent. We then have:
\begin{align}
    \textrm{Bias:} \; \mathbb{E}[\hat{H}]- \tr{H\rho} & =\sum_{\ell=1}^L\sum_{P_{\ell j}\in O_\ell}c_j(\tr{P_{\ell j}\tilde{\rho}_\ell}-\tr{P_{\ell j}\rho}), \nonumber \\ 
    \textrm{Variance:}\; \Var{\hat{H}} & =\sum_{\ell=1}^M\frac{1}{n_\ell}\Var{O_\ell},
\end{align}
where $\Var{O_\ell}=\sum_{P_{\ell j},P_{\ell k}\in O_\ell}\tr{P_{\ell k}P_{\ell j}\tilde\rho_\ell}-\tr{P_{\ell j}\tilde\rho_\ell}\tr{P_{\ell k}\tilde\rho_\ell}$.

By using Lagrange multipliers~\cite{zhu_optimizing_2024} (See~\ref{appdx:shot_lagrange}) we obtain the minimal-variance allocation $n_\ell= n\frac{\sqrt{\Var{O_\ell}}}{\sum_{j=1}^L{\sqrt{\Var{O_j}}}}$ for fixed measurement budget $n$. We can achieve a variance target $\varepsilon^2$ by assigning a total shot count $n^\varepsilon$,
$$n^\varepsilon=\frac{1}{\varepsilon^2}\left(\sum_{\ell=1}^L\sqrt{\Var{O_\ell}}\right)^2.$$

The numerator, $(\sum_{j=1}^L\sqrt{\Var{O_\ell}})^2$, is often referred to as the \emph{sample variance}~\cite{crawford_efficient_2021,yen_deterministic_2023}. The sample variance determines how many shots are necessary to reach precision target $\varepsilon$, hence grouping schemes that minimize $\sum_{j=1}^L\sqrt{\Var{O_\ell}}$ reduce the necessary shot count. 
Under ideal conditions, FC has 1.5-4$\times$ lower sampler variances, thereby requiring 50-100\% fewer measurements to reach chemical accuracy in molecular Hamiltonian estimation~\cite{yen_deterministic_2023}.  However, noisy entangling gates in FC circuits have been shown to contribute biases upwards of 50 mHartree~\cite{escudero_hardware-efficient_2023}. 

 While bias and variance are the key grouping figures of merit, we can also gain insight into the role of noisy measurements by comparing the difference between the intended ($\rho$) and actual ($\tilde\rho_\ell$) quantum states for each group. We can do so using the \emph{state fidelity} $F(\tilde\rho_\ell,\rho)\in[0,1]$. For any two quantum states $\sigma$, $\rho$, the state fidelity: \begin{equation}\label{eqn:fidelity}
    F(\sigma,\rho)=\left(\mathrm{Tr}\left[\sqrt{\sqrt{\rho}\sigma\sqrt{\rho}}\right]\right)^2,
\end{equation}
is a symmetric measure describing the similarity between two quantum states, where $F(\sigma,\rho)=1$ if and only if $\sigma=\rho$. If $\sigma=\dyad{\psi}$ is a pure state, then $F(\ket{\psi},\rho)=\ev{\rho}{\psi}$. While the bias is Hamiltonian dependent, the fidelity only depends on the trial state $\psi$ and the measurement unitary, offering a perspective into the quantum information lost in the noisy channel.

\section{Methodology}\label{sec:methods}
As introduced in the preceding section, the two primary grouping strategies (FC and QWC) present significant trade-offs in near-term quantum applications. FC measurement circuits generally incur unacceptable two-qubit gate overhead~\cite{bansingh_fidelity_2022}, while QWC circuits provide minimal measurement optimization to preserve state fidelity. Ideally, we would prefer an \emph{interpolation} between the two: a grouping strategy that retains high state fidelity while allowing entanglement for shot count optimization. An interpolation means we want to allow \emph{some} FC groups (e.g. those with lower overhead circuits), but not all.

In this section, we propose a general framework for interpolating between QWC and FC which incorporates problem context, e.g. problem or device information, when determining whether to permit a group. To make our analysis tractable, we restrict our search to functions acting over high-level device parameters. After introducing the grouping function framework (Sec.~\ref{sec:general_formulation}), we give two concrete examples. First, we consider a topology-aware grouping strategy (Sec.~\ref{sec:topology_grouping}). We then combine device topology information with a depolarizing noise model to form a commutation scheme that can adapt to different device environments, forming a Generalized backend-Aware pauLI Commutation (GALIC) grouping strategy (Sec.~\ref{sec:grouping}).

\subsection{The Hybrid Grouping Space}\label{sec:general_formulation}
The FC and QWC grouping strategies only consider the set of observables $\{P_i\}$. As many studies have shown, however, the success or failure of real quantum experiments significantly depends on external factors, such as gate and readout error~\cite{leymann_bitter_2020,mckay_benchmarking_2023}, device connectivity~\cite{leymann_bitter_2020,holmes_impact_2020}, and transpilation quality~\cite{li_tackling_2019} (to name a few). Ideally, a simultaneous measurement strategy would be able to account for performance-critical variables known prior to experimentation. For instance, hardware efficient commutativity (HEC)~\cite{miller2022hardware,escudero_hardware-efficient_2023} considers \emph{both} the target device topology and the observable set. However, it is not immediately clear how to compare nor design different grouping strategies with distinct inputs. To do so, we propose a \emph{generalized grouping function} framework for designing and analyzing simultaneous measurement strategies.

\paragraph{Grouping Functions}To generalize FC and QWC, we begin by examining generic ``grouping functions'' $f:\mathcal{P}^N\to \{0,1\}$, where $\mathcal{P}^N$ is the power set over $N$-qubit Pauli operators. That is to say, the function $f$ takes a set of Pauli strings as input and returns 1 (``permitted'') or 0 (``rejected'') to indicate whether the group is allowed.

FC and QWC can be expressed as $f_{FC}(\{P_i\})=\prod_{i\neq j} FC(P_i, P_j)$ and $f_{QWC}(\{P_i\})=\prod_{i\neq j} QWC(P_i, P_j)$ respectively: logical ANDs over purely pairwise relations. However, the design space for grouping functions $f$ is much wider. 

First, it is useful to define a way of comparing grouping functions. For convenience, we say $f_1\leq f_2$ whenever $f_1(\{P_i\})=1 $ implies $f_2(\{P_i\})=1$ for any $\{P_i\}$. In other words, $f_2$ is no more restrictive than $f_1$. We then say that $f_1=f_2$ if $f_1\leq f_2$ and $f_2\leq f_1$. Using this definition, we can define \textit{valid} grouping functions:
\begin{definition}[Grouping Function]
\label{def:grouping}
    Let $\mathcal{P}^N$ be the power set of $N$-qubit Pauli string observables. A function $f:\mathcal{P}^N\to \{0,1\}$ is a \emph{valid grouping function} if it satisfies the following two properties:
\begin{enumerate}
    \item $f\leq f_{FC}$ 
    \item $|\{P_i\}|=1 \Rightarrow f(\{P_i\})=1$
\end{enumerate}
\end{definition}

The first condition requires the Pauli strings in the set $\{P_i\}$ still mutually commute, which is a prerequisite for simultaneous measurement. The second ensures that a trivial mapping exists for all possible $\{P_i\}$, that is, we can always reduce to naive Pauli measurement.

We also make a stronger assumption that $f$ satisfies $f_{QWC}\leq f$. As previously discussed, measuring a QWC commuting group is equivalent to measuring a single Pauli string, making our assumption reasonable\footnote{One could construct a function $f$ that satisfies Requirement 2 but satisfies $f\leq f_{QWC}$, for example requiring that groups remain below a certain size, or that individual group estimator variances are below a certain tolerance. However, these are somewhat artificial constructions with little practical motivation.}. Therefore, the scheme defined by $f$ is no more restrictive than QWC.

Any $f$ satisfying our assumptions can be considered as an interpolation of QWC and FC. Like any non-parametric model, we have an infinite-dimensional space over which to optimize the bias and sample variance, as defined in Sec.~\ref{ssec:metrics}. We therefore have a \emph{generalized} grouping framework, which encapsulates previous Pauli grouping strategies as special cases.

Note that our inequality relationship $\leq$ defines a partial ordering over possible grouping functions, with $f_{QWC}$ as the least element and $f_{FC}$ as the greatest. Our framework can then have concrete implications for the analysis of grouping algorithms. $f_1\leq f_2$ implies that for any figure of merit, the optimal grouping allowed under $f_1$ can be no better than the optimal grouping allowed under $f_2$. $f_1$ can then provide a lower bound on $f_2$ performance, and conversely $f_2$ provides an upper bound on $f_1$ performance in the optimal case. 

\paragraph{Adding Context to Grouping Functions}
The efficiency of a grouping scheme can improve or degrade depending on external parameters. The specific operators being used, the trial state, device noise~\cite{bansingh_fidelity_2022}, and coupling topology can all influence the bias and variance of a simultaneous measurement estimator. Ideally, we would like to broaden our framework to incorporate external \emph{context} when testing potential simultaneous measurement groups.

We generalize $f$ to be a multivariate function over \emph{several} inputs $f(\{P_i\}, \theta_1, \theta_2...)$, where each $\theta$ is an element from an arbitrary domain.
\begin{definition}[Generalized Grouping Function] Let $\Theta_1,\Theta_2...$ be arbitrary domains. A function $f:\mathcal{P}^N\times \Theta_1\times\Theta_2\times...\to \{0,1\}$ is called a \emph{valid generalized grouping function} if, for all $\theta_1\in \Theta_1, \theta_2\in\Theta_2...$, the function $f(\cdot, \theta_1,\theta_2...):\mathcal{P}\to \{0,1\}$ satisfies the requirements of Definition~\ref{def:grouping}.
\end{definition}

We can think of each $\theta_i$ as a piece of problem context, e.g. a coupling map, noise model, user parameter, etc. By embedding context into our grouping function $f$, we can leverage knowledge of our problem or device to inform our grouping strategy.
The ordering $\leq$ between two generalized grouping functions may change depending on the context variables, as we demonstrate in the following sections.

We note that Wu et al.~\cite{wu_overlapped_2023} also introduced a unifying framework: combining grouping with tomographic techniques under the ``Overlapping Grouped Measurement'' (OGM) method. Their efforts in generalizing and unifying measurement schemes are complementary to ours. The grouping function framework proposed here is entirely compatible with the formulations used by Wu and colleagues, as generalized grouping functions simply provide another tool for constructing the sets of Pauli bases used in their estimator construction.

\begin{figure}
    \centering
    \includegraphics[width=0.9\linewidth]{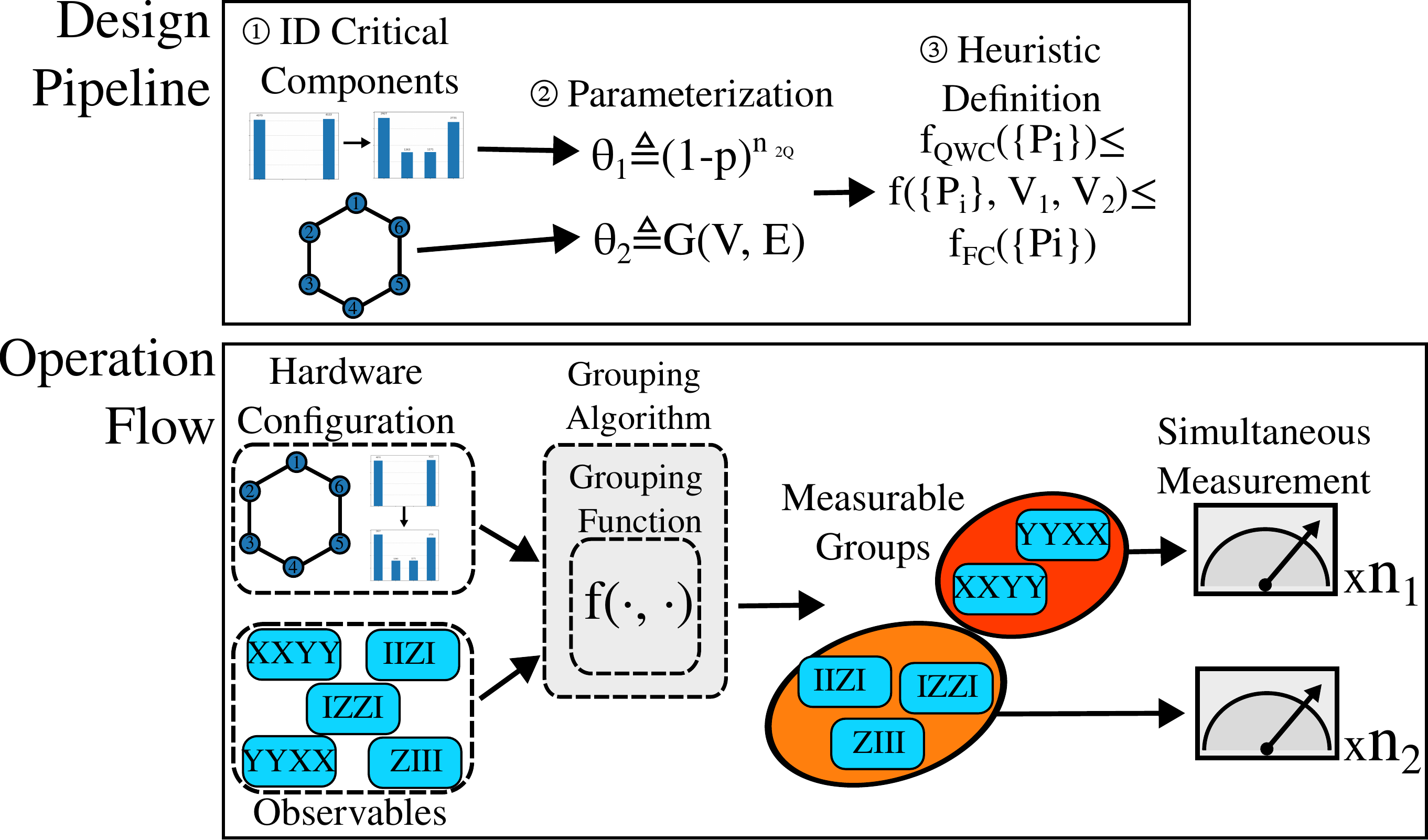}
    \caption{Illustration of the grouping function framework. [Top] The function design pipeline. Here we consider device noise and hardware topology to be ``critical components''. We then parameterize each: using a depolarizing noise model (See Sec.~\ref{sec:grouping}) and a graph data structure. We then provide a heuristic definition of a grouping function which interpolates between QWC and FC. [Bottom] An illustration of simultaneous measurement using the grouping function framework. A user-defined function $f$ takes as input a set of Pauli strings and some set of ``context variables'': here a hardware coupling graph and a noise model. The function is then used within some grouping algorithm (such as Sorted Insertion~\cite{crawford_efficient_2021}) to allow or disallow proposed groups. The grouped observables are then estimated using simultaneous measurement, allowing for reduced shot overhead.}
    \label{fig:group_func_flow}
\end{figure}

Our framework naturally defines a pipeline for efficient measurement function design, as illustrated Fig.~\ref{fig:group_func_flow}[top].
\ding{172} We begin by identifying a set of performance-critical parameters (here the device topology and noise model). \ding{173} A suitable parametrization is then found, for instance abstracting qubit connections to an undirected graph and defining a depolarizing noise model. \ding{174} Using the defined parametrization, we construct a heuristic function $f$ which interpolates between FC and QWC. Interpolation can be trivially achieved by including a fixed initial step which returns 0 if the group is not FC-compatible and 1 if the group is QWC-compatible. The remaining heuristic is designed heuristically, taking the context variables and problem application into consideration. We note that step 2 permits the possibility of variationally optimized grouping functions, if a suitable parametrization is found. However, the grouping functions defined in this work are specified in closed mathematical form rather than numerically optimized.

Fig.~\ref{fig:group_func_flow}[bottom] illustrates the simultaneous measurement pipeline defined by our framework. In a quantum experiment, the researcher provides the context variables used by the grouping function (here device noise model and topology) and the target observables. A grouping algorithm (e.g. Sorted Insertion~\cite{crawford_efficient_2021}) uses the grouping function $f$ as a kernel to allow/disallow proposed groups, ending with a collection of commuting operator sets. Each set $O_\ell=\{P_{\ell,i}\}$ is allocated a shot budget $n_\ell$ from a measurement optimization scheme~\cite{yen_deterministic_2023,zhu_optimizing_2024}, allowing for measurement-efficient quantum estimation. 

We now explore two specific applications of our framework. We first show how the HEC scheme proposed by Refs.~\cite{escudero_hardware-efficient_2023,miller2022hardware} can be naturally expressed as a generalized grouping function. Motivated by the shortcomings of HEC, we go one step further, using our framework to define a novel grouping function which considers \textit{both} gate error and device topology. By including additional function domains, our framework allows for increased function utility and flexibility.
\subsection{Grouping Function 1: Topology-Aware Hybrid Grouping}\label{sec:topology_grouping}
One of the defining features of a QPU architecture is the qubit connectivity graph. Current NISQ devices run the gamut from IBM's sparse ``heavy-hex''~\cite{ibm_link} transmon topology to fully connected trapped ion systems from IonQ~\cite{ionq_link} and Quantinuum~\cite{quantinuum_link}. 

Sparsely connected QPUs require qubit routing routines to implement entangling gates between non-connected qubits~\cite{li_tackling_2019}, incurring additional two-qubit gate overhead. Several previous works have therefore focused on various \emph{hardware-efficient} circuit designs which natively map to a device coupling graph without routing~\cite{kandala_hardware-efficient_2017}. In the same vein, we can define a {\bf hardware efficient commutation (HEC)} scheme which only requires entangling gates between connected qubits~\cite{miller2022hardware, escudero_hardware-efficient_2023}. Using our context-aware framework, we will define a grouping function $f_{HEC}$ mandating hardware-efficient measurement unitaries in the following.

Since the HEC scheme depends on device connectivity, we first formally define the coupling graph $G=(V,E)$, where $V=\{q_i\}$ represents the qubits and $E=\{(q_i, q_j)\}$ represents the set of connections between qubits. We can equivalently describe $G$ in terms of its adjacency matrix $A$, where $A_{ij}=1$ if qubits $i$ and $j$ are coupled and $0$ otherwise. Note that $A$ is symmetric with zeros on the diagonal.

From our stated requirements, any group satisfying $f_{HEC}(\{P_i\})=1$ is also fully commuting. To determine the set of Pauli strings not requiring routing, we use the concept of ``graph states''~\cite{van_den_nest_graphical_2004, Raussendorf2001, Raussendorf2003}. Graph states, which are entangled states represented by graphs ($G_s =  \{ V, E\}$), can be generated by initializing qubits to $\ket{+}$, the $+1$ eigenstate of the Pauli $X$ operator, and applying CZ gates between qubits connected by the graph edges $E$~\cite{Raussendorf2003, Hein2005}. We note that if a group of Pauli strings are stabilizers of a graph state, converting the set of Pauli strings back to Pauli $X$s acting on the qubits requires only CZ gates along the graph edges, essentially reversing the state generation~\footnote{We note that there are algorithms to most efficiently find the transformation circuit, consisting of CZ or CNOT gates along with single-qubit gates. The algorithm can be found in Ref.~\cite{crawford_efficient_2021}.}. Thus, in the HEC grouping scheme, Pauli strings $\{ P_i \}$ that stabilize a graph state $G_{\{P_i\}}$ are grouped, where $G_{\{P_i\}}$ is a sub-graph of the device's connectivity graph $G$. This ensures all diagonalization gates are hardware-native~\cite{miller2022hardware}.

We then define a generalized measurement function $f_{HEC}: \mathcal{P}\times \mathcal{G}^N\to\{0,1\}$:
\[f_{HEC}(\{P_i\}, G)=\begin{cases}
    1 & f_{FC}(\{P_i\})=1\textrm{ and }G_{\{P_i\}}\textrm{ is a subgraph of } $G$\\
    0 & \mathrm{otherwise}
\end{cases}\]
where $\mathcal{G}^N$ is the space of all possible graphs $G(V,E)$ with $|V|\geq N$. We can also express $f_{HEC}$ in pseudocode as:

\begin{algorithm}
    \begin{algorithmic}[1]
        \Procedure{$f_{HEC}$}{Pauli Set $\{P_i\}$, Coupling Graph $G$}
        \If{$\{P_i\}$ does not commute}
        \State return $0$
        \EndIf
        \State $G_{\{P_i\}}\leftarrow$ Graph State of $\{P_i\}$\Comment{E.g. using stabilizer simulations~\cite{crawford_efficient_2021}}
        \If{$G_i$ is not a subgraph of $G$}\Comment{Circuit is not device native, reject}
        \State return $0$
        \Else
        \State return $1$
        \EndIf
        \EndProcedure
    \end{algorithmic}
\end{algorithm}

A QWC set requires no entangling gates, meaning $A_z=0$, making $G_{\{P_i\}}$ a trivial subgraph of $G$. Therefore, $f_{HEC}$ satisfies $f_{QWC}\leq f_{HEC}\leq f_{GC}$, making it a valid hybrid scheme within our framework.

\subsection{Grouping Function 2: Noise-Aware Hybrid Grouping}\label{sec:grouping}
Unfortunately, topology-only grouping functions are insufficient to control biases across device models. $f_{HEC}$ reduces to $f_{FC}$ on fully-connected trapped-ion devices despite gate errors potentially exceeding 3\%~\cite{pelofske_high-round_2023}. Moreover, two-qubit gate costs can vary significantly between devices with the same coupling graph, a factor not considered by $f_{HEC}$. IBM Eagle and Heron processors, for instance, have been shown to differ in 2-qubit gate fidelity by $>2\times$~\cite{mckay_benchmarking_2023} despite sharing a heavy-hex topology~\cite{ibm_link}. Lower error devices may permit limited qubit routing, making HEC overly restrictive, particularly in future architectures.

Hardware topology can provide the \emph{number} of CNOT gates needed to diagonalize a Pauli group, however it does not account for the fidelity cost \emph{per} CNOT gate. Therefore, HEC cannot construct an accurate model of simultaneous measurement cost, potentially leading to unacceptable measurement circuit error. To rectify this, we can consider both device fidelity and topology to construct a more complete picture of Pauli group error.

Fortunately, many hardware providers report estimated gate error, allowing us to define a generalized grouping function that combines device fidelity and topology. In this sub-section we use our design pipeline to formulate a ``backend-aware'' grouping function: capable of adapting to device models with varying connectivity and fidelity. Having identified gate error and topology as critical parameters, we begin by developing a depolarizing noise model parameterized by two-qubit gate error $p$ and a target accuracy $\epsilon_{\text{tar}}$. Using our parameterized models, we construct a valid generalized grouping function which can adapt to different qubit connectivity graphs and noise environments: the Generalized backend-Aware pauLI Commutativity (GALIC) scheme. 

\paragraph{Expectation Values with Finite Quantum Noise}
Since our goal is to control grouping-derived estimator error, we require a model of estimator bias in the presence of quantum noise. We assume that device infidelity can be reasonably approximated by a depolarizing noise channel $\mathcal{E}$. The effect of $\mathcal{E}$ on $N$-qubit quantum state $\rho$ is given by:
\begin{equation}
    \mathcal{E}(\rho)=(1-p_\text{err})\rho+p_\text{err} \frac{I}{2^N},
\end{equation}
where $p_\text{err}$ is the probability of depolarizing faults occurring. We further assume that the total gate error probability in the measurement circuit contributes to the error probability of a global depolarization channel, i.e., $p_\text{err} = 1 - (1-p)^{n_\text{2q}}$, where $n_\text{2q}$ is the total number of 2-qubit gates and $p$ is the median 2-qubit gate error probability reported by hardware providers.

Depolarizing noise models are insufficient for complete noise channel characterization, however, they can serve as reasonable approximations of device behavior~\cite{georgopoulos_modeling_2021}, particularly for superconducting qubits~\cite{krantz_quantum_2019}. Moreover, depolarizing models have been effectively employed in error-mitigation protocols~\cite{he_zero-noise_2020,temme_error_2017,vovrosh_simple_2021}, motivating their use for reducing estimator biases in Pauli grouping schemes. We neglect the impact of single-qubit gates, since entangling gate noise is several factors greater and typically dominates~\cite{wei_hamiltonian_2022,arute_quantum_2019,arute_supplementary_2019}. By formulating our noise model in terms of two-qubit gates, we can estimate the impact of entangling operations on expectation estimates and develop a grouping function which limits two-qubit gate counts accordingly.

After applying the noise channel, the expectation value of estimator $\hat P$ is given by:
\begin{equation}
    \label{eqn:h_expect}
    \mathbb{E}[\hat P]=\mathrm{Tr}[\mathcal{E}(\rho) P]= (1-p_\text{err}) \mathrm{Tr}[\rho P]+ p_\text{err} \frac{\mathrm{Tr}(P)}{2^N} =(1-p)^{n_{2q}}\mathrm{Tr}[P\rho],
\end{equation}
where we use the fact that Pauli strings are traceless, and hence $\tr{P} = 0$.

With some algebraic manipulation, we obtain:
\begin{equation}
    1+\frac{\mathbb{E}[\hat P]-\mathrm{Tr}[P\rho]}{\mathrm{Tr}[P\rho]}=(1-p)^{n_\text{2q}}
\end{equation}

Note that by Eqn.~\eqref{eqn:h_expect}, we have $|\mathbb{E}[\hat P]|\leq |\mathrm{Tr}[P\rho]|$ and we assume $\mathrm{sign}(\mathbb{E}[\hat P])=\mathrm{sign}(\mathrm{Tr}[P\rho])$. Therefore we have $(\mathbb{E}[\hat P]-\mathrm{Tr}[P\rho])/\mathrm{Tr}[P\rho] = -|\mathbb{E}[\hat P]-\mathrm{Tr}[P\rho]|/|\mathrm{Tr}[P\rho]|$. 
By defining the relative error $\epsilon\triangleq\frac{|\mathbb{E}[\hat P]-\mathrm{Tr}[P\rho]|}{|\mathrm{Tr}[P\rho]|}$,
\begin{equation}
\begin{split}\label{eqn:bias}
    \epsilon&=1-(1-p)^{n_\text{2q}}\\
\end{split}
\end{equation}

For $p>0$, we have $\epsilon>0$, meaning an asymptotically \emph{biased} estimator. Error mitigation techniques such as zero-noise extrapolation~\cite{temme_error_2017,he_zero-noise_2020} or subspace expansion~\cite{mcclean_decoding_2020} can correct for biased estimates at the cost of additional sample overhead: \emph{precisely what simultaneous measurement schemes are supposed to avoid}.

\paragraph{Developing a Noise-Informed Grouping Function}
However, we can use our noise model to restrict the bias introduced by Pauli measurement circuits. Inverting the bound, we obtain:
\begin{equation}
    \frac{\log(1-\epsilon)}{\log(1-p)}=n_\text{2q}.
    \label{eqn:ratio}
\end{equation}

For fixed error tolerance $\epsilon_\text{tar}$ and backend-provided gate error probability $p$, we can use Eqn.~\eqref{eqn:ratio} to bound the number of two-qubit gates allowed for each measurement circuit, i.e.,

\begin{equation}\label{eqn:2qlim}
    n_\text{2q}\leq\frac{\log(1-\epsilon_\text{tar})}{\log(1-p)}.
\end{equation}

However, exactly knowing $n_{2q}$ would require both constructing and transpiling the measurement circuit for every \emph{proposed} group. Grouping heuristics typically scale $O(M N)$ to $O(M^2 N)$~\cite{crawford_efficient_2021,gokhale_optimization_2020,gokhale2020n} for $M$ Paulis and $N$ qubits, with circuit construction and routing passes taking $O(N^3)$~\cite{crawford_efficient_2021} and $O(N^{2.5})$~\cite{li_tackling_2019} respectively. The overall cost of Pauli grouping would then exceed $O(N^4M)\simeq O(N^{8})$, as $M$ scales $O(N^4)$ in second-quantized Hamiltonians~\cite{tilly_variational_2022}. 

An alternative strategy is to \textbf{upper bound} the number of two-qubit gates using circuit construction properties. Recall that the Gaussian elimination algorithms used to construct measurement unitaries use at most $\frac{1}{2}N_{AC}(\{P_i\})(N_{AC}(\{P_i\})-1)$ entangling gates, where $N_{AC}(\{P_i\})$ is the number of anti-commuting qubits in $\{P_i\}$. We thereby obtain an upper bound the number of CNOT gates in the \emph{untranspiled} measurement circuit.

During transpilation, CNOT gates between distant qubits are implemented via SWAP-based routing. For a pair of qubits separated by a distance $D$ in the device connection graph, routing requires $(D-1)$ SWAP gates to implement the entangling operations. Each SWAP gate can be decomposed as 3 CNOT gates~\cite{nielsen_quantum_2011}, meaning a total CNOT overhead of $3(D-1)$ gates. Given the device topology, we can conservatively bound the total CNOT count as
$[3(D_\text{max}(G,\{P_i\})-1) + 1]N_{AC}(\{P_i\})(N_{AC}(\{P_i\})-1)/2$, where $D_\text{max}(G,\{P_i\})$ is the maximum number of graph hops between two anti-commuting qubits in $\{P_i\}$. We can efficiently find the shortest path lengths between all qubits in $O(|E|N+N^2)$ time using breadth-first search~\cite{cormen2022introduction}: a step which only needs to be performed once for any coupling graph. We therefore reduce the complexity back to the original $O(M N)-O(M^2 N)$. grouping algorithm complexity.

Given the expected error tolerance $\epsilon_\text{tar}$, the device  coupling graph $G$ and gate error probability $p$, we define an upper bound on $N_{AC}(\{P_i\})$:
\begin{equation}
\label{eqn:galic_condition}
    \overbrace{\frac{1}{2}N_{AC}(\{P_i\}) (N_{AC}(\{P_i\}) - 1)}^{\textrm{Max. Logical CNOTs}}\cdot \overbrace{(3(D_\text{max}(G,\{P_i\})-1) + 1)}^{\textrm{Max. Routing Overhead}} \leq \frac{\log(1-\epsilon_\text{tar})}{\log(1-p)}.
\end{equation}

We can then define the {\bf Generalized backend-Aware pauLI Commutation (GALIC)} scheme with grouping function $f_{GALIC}:\mathcal{P}\times \mathbb{R}\times\mathbb{R}\times \mathcal{G}^N\to \{0,1\}$:
$$f_{GALIC}(\{P_i\}, p, \epsilon, G)=\begin{cases}
    1 & f_{FC}(\{P_i\})=1 \ \textrm{and}\;N_{AC}(\{P_i\}), \;D_\text{max}(G,\{P_i\})\textrm{ satisfy Eqn.\eqref{eqn:galic_condition}}\\
    0 & \textrm{otherwise}
\end{cases}$$
or, expressed in pseudocode:
\begin{algorithm}
    \begin{algorithmic}[1]
        \Procedure{$f_{GALIC}$}{Pauli Set $\{P_i\}$, 2Q Error $p$, Target $\epsilon_{\text{tar}}$, Coupling Graph $G$}
        \If{$\{P_i\}$ does not commute}
        \State return $0$
        \EndIf
        \State $\{q_i\}\leftarrow$ Set of Anti-Commuting Qubits in $\{P_i\}$
        
        \State $N_{AC}\leftarrow |\{q_i\}|$\Comment{Number of anti-commuting qubits}
        \State $D_{\max}\leftarrow$ Maximum Distance Between $\{q_i\}$ on $G$
        \If{$N_{AC},\;D_{\max},\;p\;,\epsilon_{\text{tar}}$ satisfy Eqn.~\eqref{eqn:galic_condition}}
        \State return $1$\Comment{Acceptable 2-qubit overhead}
        \Else
        \State return $0$\Comment{Too many potential 2-qubit gates$\to$ reject}
        \EndIf
        \EndProcedure
    \end{algorithmic}
\end{algorithm}

By incorporating both device noise and topology, $f_{GALIC}$ can effectively adapt to different backends while mitigating estimator accuracy losses from measurement entanglement. It also provided us a tool to further explore the design space between QWC and FC grouping schemes, and better understand the design of measurement schemes in the presence of device noise and topology. 

Before evaluating GALIC, we consider the relationships between different grouping functions using our partial ordering. Since both $f_{HEC}$ and $f_{GALIC}$ are context aware, their relative ordering depends on the noise level and topology of the target device and the target error $\epsilon_{tar}$. 

\begin{table}[h]
\caption{Relation between the GALIC grouping scheme and QWC, HEC, and FC schemes.}
\label{tab:cases}
\begin{indented}
    \item[]\begin{tabular}{@{}r||c|c}
    \br
        &  Sparse $G$ & Complete $G$\\
        \mr
        $p>\epsilon_{tar}$ (High Noise) & $f_{QWC}=f_{GALIC}\leq f_{HEC}\leq f_{FC}$ &$f_{QWC}=f_{GALIC}\leq f_{HEC}= f_{FC}$\\
        $p\leq\epsilon_{tar}$ (Low Noise)  & $f_{QWC}\leq f_{GALIC}\;?\;f_{HEC}\leq f_{FC}$ & $f_{QWC}\leq f_{GALIC}\leq f_{HEC}= f_{FC}$\\
        $\lim p/\epsilon_{tar}\to 0$ (Ideal)  & $f_{QWC}\leq f_{HEC}\leq f_{GALIC}= f_{FC}$ & $f_{QWC}\leq f_{GALIC}= f_{HEC}= f_{FC}$\\
    \br
    \end{tabular}
    \label{tab:relations}
\end{indented}
\end{table}

All relationships in Tab.~\ref{tab:relations} depend on local properties (coupling graph connectivity and individual gate errors). If the connectivity does not increase with system size and the gate fidelity remains constant, then the partial order will persist in the large $N$ limit.

Whenever $G$ is complete, $f_{HEC}$ reduces to $f_{FC}$ regardless of noise. Likewise, $f_{GALIC}$ reduces to $f_{FC}$ whenever machine error becomes negligible. All entangling schemes become equivalent for complete architectures with negligible error. 

In the case where $p\leq \epsilon_\text{tar}$, we cannot determine a strict relationship between $f_{GALIC}$ and $f_{HEC}$, illustrating the \emph{partial} ordering within our framework. If device noise is high, GALIC will likely be more restrictive than HEC (permitting fewer CNOTs). If device noise sufficiently low, then GALIC will be more lenient, allowing for non-native entangling operations forbidden under HEC. In current-generation NISQ devices, GALIC is generally more restrictive than HEC, as we numerically show in Sec.~\ref{sec:eval}. In the near term, our simulated results show that restrictive grouping is necessary to maintain state fidelity and control estimator biases. However, as gate fidelity improves, GALIC measurement efficiency will approach FC. The joint consideration of topology and noise allows us to compare the impact of quantum design space on estimator performance in Sec.~\ref{sec:discuss}.

\section{Evaluation}\label{sec:eval}
In this section, we compare GALIC and HEC with QWC and FC across several device models and molecular Hamiltonians with a focus on state fidelity, estimator bias, and sampling error.

\subsection{Experimental Methodology}
For all grouping schemes, we use the Sorted Insertion (SI) heuristic to perform operator grouping, which was reported by Ref.~\cite{crawford_efficient_2021}. SI sorts Pauli operators by coefficient magnitude, then greedily constructs commuting sets in descending order. Despite its simplicity, SI has been noted for its high performance and low time complexity, and therefore has been used as the baseline grouping algorithm in numerous recent works~\cite{yen_deterministic_2023,miller2022hardware,choi_improving_2022,bansingh_fidelity_2022}. We provide full pseudocode and implementation details in~\ref{appdx:si}.

\begin{table}[h]
    \caption{Parameters used for device noise model construction. $T_{1}$ and $T_2$ are the longitudinal and transversal relaxation times of the qubits, $p_{1Q}$ ($p_{2Q}$) and $t_{1Q}$ ($t_{2Q}$) are the depolarizing error probability and time duration of one-qubit (two-qubit) gates, respectively.}
    \label{tab:noise_parameters}
    \begin{indented}
    \item[]\begin{tabular}{@{}l|cc|cc}
    \br
        Provider & \multicolumn{2}{c}{\textbf{IBM}} & \multicolumn{2}{c}{\textbf{IonQ}}\\\midrule
        Parameter & Sherbrooke & Torino& Aria1& Forte\\
        \midrule
        $p_{1Q}$ (\%) & 0.02 & 0.03 & 0.06 & 0.02\\
        $p_{2Q}$ (\%) &  0.7  & 0.3 & 8.6 & 1.0\\
        $T_1$ ($\mu s$) & 259.7 & 160.5 & \SI{100e6}{} & \SI{100e6}{}\\
        $T_2$ ($\mu s$)& 182.3 & 122.4 & \SI{1e6}{}\ & \SI{1e6}{}\\
        $t_{1Q}$ ($\mu s$)& 0.057 & 0.032 & 135 & 130\\
        $t_{2Q}$ ($\mu s$)& 0.533 & 0.068 & 600 & 970\\
        \br
    \end{tabular}
    \end{indented}
 \end{table}

Numerical tests used NWQ-Sim~\cite{wang2023enabling,li_density_2020,li_sv-sim_2021} density matrix simulations to model device behavior and Qiskit utilities for noiseless simulation and operator diagonalization~\cite{qiskit2024}. Noise models were constructed by composing depolarizing and thermal relaxation noise channels for both single and two-qubit gates with NWQ-Sim~\cite{wang2023enabling,li_density_2020,li_sv-sim_2021} utilities, using mean noise parameters taken from IBM and IonQ device information as our model inputs, shown in Table~\ref{tab:noise_parameters}. We model the sparse Sherbrooke and Torino devices from IBM~\cite{ibm_quantum_resources} and the fully-connected Aria1 and Forte devices from IonQ~\cite{ionq_link} \footnote{The noise parameters, taken from provider randomized benchmarking~\cite{knill2008randomized} data in mid-June 2024, are liable to fluctuate. Therefore our reported noise levels may differ from those reported elsewhere~\cite{pelofske_high-round_2023} or current device values}.

Our testing computes exact biases and variances using output density matrices, hence we do not model readout error (as no readout occurs). 
We test each measurement scheme on 5 molecular Hamiltonians, summarized in Tab.~\ref{tab:hamiltonians}. Each Hamiltonian is constructed using the minimal STO-3G basis set using PySCF~\cite{sun2018pyscf} with diatomic bond lengths of 1 Å. Due to the overhead of density matrix simulation, we limit our testing to 7 orbital (14 qubit) problems. 
\begin{table}[ht]
    \caption{STO-3G Molecular Hamiltonians used for testing. Each molecule has inter-atomic bond lengths of 1 Å. We report the number of electrons ($N_e$), the number of spatial orbitals ($N_{Orbs}$), the number of Fermionic operator terms in the Hamiltonian ($N_{Ops}$), the number of qubits ($N_Q$), and the number of Pauli strings ($M$) following the Jordan-Wigner transformation.}
    \label{tab:hamiltonians}
\begin{indented}
    \item[]\begin{tabular}{@{}c c c c c c c}
    \br
        Molecule & Geometry & $N_{e}$ & $N_{Orbs}$ & $N_{Ops}$ & $N_Q$ & $M$\\
        \mr
        $\mathrm{H_4}$ & Square & 4 & 4 &524 & 8& 121\\ 
        $\mathrm{LiH}$ & - & 4 & 6 & 1860 & 12&631\\ 
        $\mathrm{H_6}$ & Hexagonal & 6 & 6 &2092& 12& 703\\ 
        $\mathrm{BeH_2}$ & Chain & 6 & 7 & 3150 & 14 & 1086 \\ 
        $\mathrm{H_2O}$ & 107\textdegree & 10 & 7 & 1938 & 14 & 666\\ 
        \br
    \end{tabular}
\end{indented}
\end{table}
For NISQ device validation, we estimate $\mathrm{H_4}$ expectation values on the IBM Kyoto system with shots allocated to the minimal-variance allocation strategy~\cite{miller2022hardware,crawford_efficient_2021,zhu_optimizing_2024} (See Appendix~\ref{appdx:shot_lagrange}). Variances for shot allocation are computed exactly, however frameworks exist to estimate $\mathrm{Var}[O_i]$ for classically intractable system sizes~\cite{wu_overlapped_2023, yen_deterministic_2023}.

IBM coupling graphs are generated as distance 3 heavy-hex graphs, while IonQ topologies are simply fully connected. For design space testing in Sec.~\ref{sec:discuss}, we use NetworkX~\cite{hagberg2008exploring} to generate random regular coupling graphs with fixed coupling degree.

Data to reproduce plots and tables can be found at Ref.~\cite{matt_x_burns_2024_13621729}.
\subsection{Numerical Experiments}\label{sec:results}
\subsubsection{State Fidelity}
\begin{figure}[h!]
    \centering
    \includegraphics[width=1\linewidth]{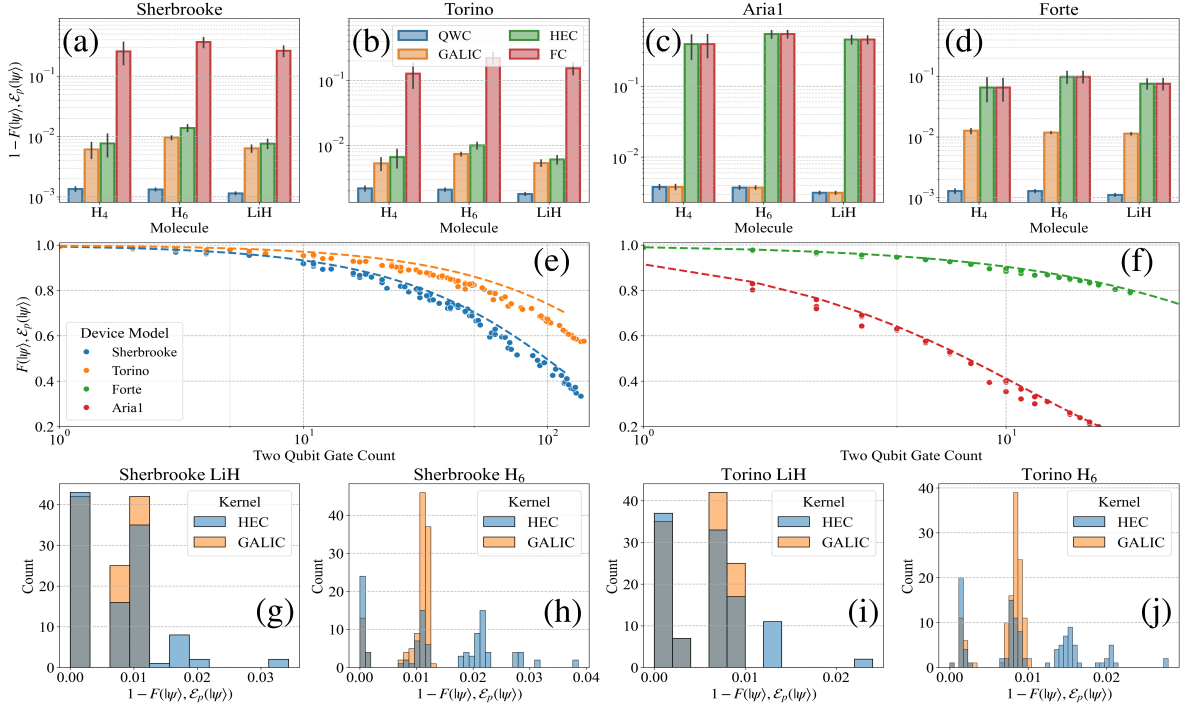}
    \caption{(a-d) Fidelity distances from the ideal state for each measurement scheme on three molecular Hamiltonians across each device model. (e,f) The scaling of fidelity with increasing CNOT count. Dashed lines show the expected scaling from Eqn.~\eqref{eqn:fidelity_bound}. (g-j)  Infidelity distribution for GALIC and HEC on the sparse IBM device models for 12 qubit problems.}
    \label{fig:fidelity}
\end{figure}

To illustrate the effects of noise and limited quantum hardware connectivity, we compare the quantum state before and after applying measurement circuits across various measurement protocols. The state prior to the measurement circuit is the noiseless pure state $\ket{\psi}$, while the post-measurement state is denoted as $\tilde{\rho}$. Since the measurement circuits vary depending on the set of Pauli strings being measured, we calculate the average fidelity across different observable groups for each measurement scheme. The resulting average infidelities are presented in Fig.~\ref{fig:fidelity}.

As shown in Fig.~\ref{fig:fidelity}(a) to~(d), QWC consistently maintained the lowest infidelity, remaining below 0.1 in each case. FC infidelity consistently exceeded 0.5, indicating a significant loss in quantum information from errors in the measurement circuit. On the sparsely connected Sherbrooke and IBM topologies, GALIC and HEC interpolated between the two extremes with approximately matched means. However, the average variance in GALIC infidelity across the three Hamiltonians was 3.5$\times$ lower than HEC. The latter exhibits distinctly long tails in its infidelity distribution, shown in Fig.~\ref{fig:fidelity}(g) to~(j). HEC also contained more singleton Paulis for the $H_6$ problem, causing the low-infidelity spike. However, this was not necessarily true for all molecules.

The difference between GALIC and HEC becomes more pronounced in the fully connected models [Forte and Aria1, see Fig.~\ref{fig:fidelity}(c), (d)]. Here HEC becomes equivalent to FC, even with the high 2-qubit gate errors of the Aria1 backend (8.5\%). Consequently, the average infidelity exceeds 0.39 for all Aria1 groups and 0.06 for all Forte groups. For Aria1, GALIC reduces to QWC to minimize state degradation. However, the Forte model permits one entangling gate per GALIC group, maintaining average infidelities below 0.015 while relaxing the QWC scheme.

We then check how well our depolarization noise model can predict the performance of quantum devices with thermal relaxations. Using the depolarizing noise model from Sec.~\ref{sec:grouping}, we can approximately bound the fidelity in terms of the number of CNOT gates ($n_\text{cx}$). For a depolarizing channel $\mathcal{E}_p$ applied to pure state $\dyad{\psi}$, the fidelity loss can be characterized as:
\begin{equation}\label{eqn:fidelity_bound}
\begin{split}
    F(\ket{\psi},\mathcal{E}_p(\ket{\psi}))&=\bra{\psi} \left[ (1-p)^{n_\text{cx}}\dyad{\psi}+\frac{(1-(1-p)^{n_\text{cx}})}{2^N}I\right]| \ket{\psi}\\
    &= \frac{1}{2^N} + \left( 1 -\frac{1}{2^N}\right) \left( 1 - p\right)^{n_\text{cx}},
\end{split}
\end{equation}
where $N$ is the number of qubits. In the limit $n_\text{cx}\to\infty$ with finite $p$, $F(\ket{\psi},\mathcal{E}_p(\ket{\psi}))\to 2^{-N}$ as $\mathcal{E}_p(\ket{\psi})$ approaches a uniform mixed state.

Since there are other sources of noise in the system, especially when the qubits are idling, waiting for gate implementation on other qubits, the qubits still suffer thermal relaxation noise, we expect Eqn.~\eqref{eqn:fidelity_bound} to serve as an upper bound on the true fidelity. Fig.~\ref{fig:fidelity}e and~f shows the observed scaling of state fidelity with increasing two-qubit gate counts. Markers indicate simulated results, while the lines show the expected scaling using Eqn.~\eqref{eqn:fidelity_bound}. Though with small quantitative differences, in all cases, Eqn.~\eqref{eqn:fidelity_bound} gives the correct qualitative behavior.

\subsubsection{Bias and Variance}

\newcommand{\red}[1]{{\color{red} #1}}
\begin{table}[h!]
    \caption{Bias and sample variance of the estimated molecular ground state energies. The exact ground state energy is solved by direct diagonalization of the Hamiltonian matrix. Biases exceeding chemical accuracy (1 kcal/mol) are shown in \red{red}, with summary statistics at the bottom.}
    \label{tab:bias_var}
    \centering
\scalebox{0.85}{\begin{tabular}{ll|llll|rrrr}
\toprule
 &  & \multicolumn{4}{c}{\textbf{Bias (kcal/mol)}} & \multicolumn{4}{c}{\textbf{Sample Variance (kcal$^2$/mol$^2$)}} \\
 & Kernel & QWC & GALIC & HEC & FC & QWC & GALIC & HEC & FC \\
Molecule & Target &  &  &  &  &  &  &  &  \\
\midrule
\multirow[t]{4}{*}{$\mathrm{H_4}$} & Aria1 & 0.696 & 0.696 & \red{101.235} & \red{101.235} & 7.28e+05 & 7.28e+05 & 3.27e+05 & 3.27e+05 \\
 & Forte & 0.197 & \red{2.336} & \red{18.737} & \red{18.737} & 7.28e+05 & 5.46e+05 & 2.96e+05 & 2.96e+05 \\
 & Sherbrooke & 0.228 & \red{1.170} & \red{1.361} & \red{74.880} & 7.28e+05 & 6.10e+05 & 5.88e+05 & 3.20e+05 \\
 & Torino & 0.407 & 0.962 & \red{1.173} & \red{38.931} & 7.29e+05 & 6.09e+05 & 5.87e+05 & 3.18e+05 \\
\cline{1-10}
\multirow[t]{4}{*}{$\mathrm{H_6}$} & Aria1 & 0.448 & 0.448 & \red{70.355} & \red{70.355} & 5.49e+06 & 5.49e+06 & 1.08e+06 & 1.08e+06 \\
 & Forte & 0.127 & \red{1.343} & \red{15.759} & \red{15.759} & 5.61e+06 & 4.48e+06 & 1.24e+06 & 1.24e+06 \\
 & Sherbrooke & 0.147 & \red{1.126} & \red{1.742} & \red{55.028} & 6.19e+06 & 4.22e+06 & 3.58e+06 & 1.06e+06 \\
 & Torino & 0.262 & 0.872 & \red{1.430} & \red{32.958} & 5.56e+06 & 4.22e+06 & 3.47e+06 & 1.08e+06 \\
\cline{1-10}
\multirow[t]{4}{*}{$\mathrm{LiH}$} & Aria1 & 0.093 & 0.093 & \red{11.459} & \red{11.459} & 5.57e+05 & 5.57e+05 & 3.57e+05 & 3.57e+05 \\
 & Forte & 0.026 & 0.316 & \red{1.716} & \red{1.716} & 5.57e+05 & 8.80e+05 & 3.46e+05 & 3.46e+05 \\
 & Sherbrooke & 0.031 & 0.266 & 0.378 & \red{5.329} & 5.53e+05 & 5.38e+05 & 4.89e+05 & 3.42e+05 \\
 & Torino & 0.054 & 0.215 & 0.291 & \red{2.743} & 5.52e+05 & 5.38e+05 & 4.88e+05 & 3.38e+05 \\
\cline{1-10}
\multirow[t]{4}{*}{$\mathrm{H_2O}$} & Aria1 & 0.321 & 0.321 & \red{37.464} & \red{37.464} & 1.36e+07 & 1.36e+07 & 3.43e+06 & 3.43e+06 \\
 & Forte & 0.091 & \red{1.064} & \red{6.872} & \red{6.872} & 1.39e+07 & 9.75e+06 & 3.42e+06 & 3.42e+06 \\
 & Sherbrooke & 0.105 & 0.763 & \red{1.022} & \red{24.824} & 1.36e+07 & 1.07e+07 & 1.08e+07 & 3.42e+06 \\
 & Torino & 0.187 & 0.652 & 0.938 & \red{15.055} & 1.39e+07 & 1.07e+07 & 1.08e+07 & 3.42e+06 \\
\cline{1-10}
\multirow[t]{4}{*}{$\mathrm{BeH_2}$} & Aria1 & 0.146 & 0.146 & \red{17.004} & \red{17.004} & 1.78e+06 & 1.78e+06 & 4.12e+05 & 4.12e+05 \\
 & Forte & 0.041 & 0.449 & \red{2.765} & \red{2.765} & 1.78e+06 & 1.71e+06 & 4.06e+05 & 4.06e+05 \\
 & Sherbrooke & 0.048 & 0.369 & 0.518 & \red{12.803} & 1.82e+06 & 1.32e+06 & 9.52e+05 & 4.07e+05 \\
 & Torino & 0.086 & 0.303 & 0.394 & \red{6.944} & 1.79e+06 & 1.32e+06 & 9.79e+05 & 4.07e+05 \\
\cline{1-10}

\midrule
& Summary & \multicolumn{4}{c}{\# Exceeded 1 kcal/mol} & \multicolumn{4}{c}{Mean Reduction vs. QWC (IBM/IonQ)}\\\cline{1-10}
& & 0 & 5 & 15 & 20 & 1/1 & 1.26/1.07 & 1.33/3.42 & 3.57/3.42\\\cline{1-10}
\bottomrule
\end{tabular}}
\end{table}

In this subsection, we compare the performance of GALIC, HEC, FC, and QWC grouping schemes, focusing on sample variance and measurement circuit biases. Using the molecular ground energy problem as a study case, To isolate the effects of device noise and connectivity on the measurement circuits from the state preparation process, we first obtain the exact ground state wavefunction through diagonalization. From this state, we apply the respective grouping schemes, and extract the molecular Hamiltonian's expectation value from the density matrices resulting from noisy measurement circuits.
The energy bias with respect to the true ground state energy and the sample variances for each grouping scheme under various device noise models are reported in Table~\ref{tab:bias_var}.
Biases exceeding chemical accuracy (CA) of 1 kcal/mol are highlighted in \red{red}. 

While FC grouping led to the lowest variances (mean 3.4$\times$ reduction versus QWC), biases resulting from entanglement-heavy unitaries exceeded chemical accuracy in 100\% of the problems/devices tested. In contrast, QWC never exceeded chemical accuracy, paying for its high fidelity in increased variance. In the extreme case of $\mathrm{H_2O}$, QWC variance exceeded $\SI{1.39e7}{kcal^2/mol^2}$(34 Hartree$^2$).

HEC and GALIC both interpolated between these extremes. While HEC generally achieved lower biases than FC, it was still biased above CA in 15 out of 20 cases. In contrast, GALIC only exceeded CA in 5 cases, and never by more than 1 kcal/mol. We observe that the highly correlated $H_4$ and $H_6$ systems exacerbate the loss in state fidelity for all simultaneous measurement schemes, including QWC.

\begin{figure}[t!]
    \centering
    \includegraphics[width=1\linewidth]{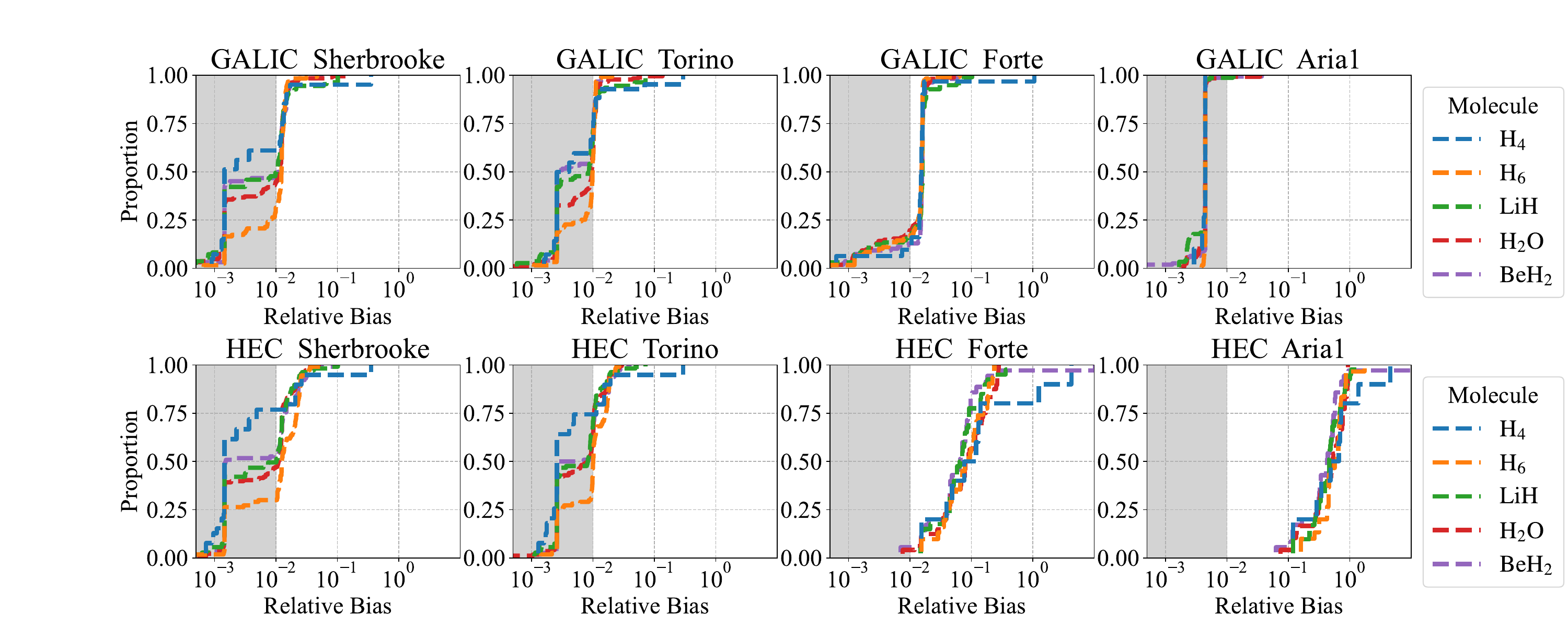}
    \caption{Cumulative distribution plots for the relative bias $\epsilon_\text{tar}$ (Eqn.~\eqref{eqn:bias}) for each observable group estimator $\hat O_i$. The x-axis denotes the relative bias magnitude, the y-axis shows the proportion of observable groups with bias less than or equal to a given relative bias. Note in the case of Forte and Torino, the GALIC relative bias approximately asymptotes at the target value $\SI{1e-2}{}$.}
    \label{fig:bias_ecdf}
\end{figure}

Despite having the same coupling architecture, the Sherbrooke device model resulted in $\sim$20\% larger biases than Torino. The Sherbrooke system has $\sim$ 8x longer gate durations than Torino despite having comparable $T_1/T_2$ times. We then conclude that thermal error was a much more critical issue in the Sherbrooke noise channel. The Forte model also exhibited larger biases, despite having low thermal error. This can be attributed to the presence of noisy single qubit gates in the measurement circuit, which contributed their own finite depolarization error. Our device model presupposes that two-qubit depolarizing error is the dominant noise source in the system, hence our simplified heuristic performs less effectively. The differences between device models can be more clearly seen by comparing the cumulative distribution of relative biases, shown for GALIC and HEC in Fig.~\ref{fig:bias_ecdf}. 

Recall that the GALIC relative bias target was $\epsilon_\text{tar}\leq\SI{1e-2}{}$. GALIC approximately met the target on the Forte, Sherbrooke, and Torino models, showing a clear jump near $\SI{1e-2}{}$. However, the specific ``jump'' point varied depending on the accuracy of the depolarizing channel in describing device behavior. The Forte, Sherbrooke, and Torino distributions each reach the 90\% threshold at $\SI{1.7e-2}{}$, $\SI{1.49e-2}{}$, and $\SI{1.13e-2}{}$ respectively, further indicating that the two-qubit depolarizing error model was not sufficient to completely describe channel behavior and precisely bound the bias, particularly for the Forte and Sherbrooke devices. However, as gate times decrease and/or gate fidelities increase, as in the Torino system, we expect that our depolarization-based heuristic will become a better approximation of device behavior. The relative biases are also much more controlled and consistent compared to the HEC results. HEC distributions reach 90\% on the Forte, Sherbrooke and Torino models at $\SI{2.32e-1}{}$,  $\SI{2.4e-2}{}$, and $\SI{1.8e-2}{}$ respectively: up to 13.6$\times$ larger than GALIC.

Table~\ref{tab:circ_sim} shows average measurement circuit characteristics for the sparse IBM Torino and dense IonQ Forte models. By removing/penalizing routing, both GALIC and HEC provide substantially shallower circuits than FC on the Torino topology. However, GALIC measurement circuits use $30\%$ fewer entangling gates than HEC unitaries on average, and up to $4\times$ less in the maximal case. Despite the CNOT reduction, GALIC only increases the number of commuting groups by 4\% on average.

\begin{table}[]
\caption{Average measurement circuit characteristics for entanglement grouping schemes in numerical testing for Torino and Forte backends. Note the differences in group counts for FC are due to slight fluctuations in the STO-3G basis coefficients from run-to-run which impacted the Sorted Insertion ordering for low-magnitude Pauli terms. Estimator performance, however, was found to be robust to these variations.}
    \label{tab:circ_sim}
    \centering
\scalebox{0.82}{\begin{tabular}{ll|rrrr|lll|lll}
\toprule
 &  & \multicolumn{4}{c}{\textbf{Groups}} & \multicolumn{3}{c}{$\bm{N_{2Q}}$ (Mean/Max)} & \multicolumn{3}{c}{\textbf{Depth} (Mean/Max)} \\
Target & Molecule & QWC & GALIC & HEC & FC & GALIC & HEC & FC & GALIC & HEC & FC \\
\midrule
\multirow[t]{5}{*}{Forte} 
 & $\mathrm{H_4}$ & 49 & 30 & 8 & 8 & 0.9/1 & 5.5/12 & 5.5/12 & 11.3/12 & 23.9/49 & 23.9/49 \\
 & $\mathrm{H_6}$ & 191 & 126 & 32 & 32 & 1.0/1 & 9.9/22 & 9.9/22 & 11.5/12 & 30.7/58 & 30.7/58 \\
 & $\mathrm{LiH}$ & 143 & 96 & 38 & 38 & 1.0/1 & 8.3/24 & 8.3/24 & 11.5/12 & 29.4/62 & 29.4/62 \\
 & $\mathrm{BeH_2}$ & 157 & 107 & 30 & 30 & 1.0/1 & 11.5/32 & 11.5/32 & 11.5/12 & 34.2/66 & 34.2/66 \\
 & $\mathrm{H_2O}$ & 246 & 161 & 48 & 48 & 1.0/1 & 12.0/30 & 12.0/30 & 11.6/12 & 35.3/66 & 35.3/66 \\
\cline{1-12}
\multirow[t]{5}{*}{Torino} 
 & $\mathrm{H_4}$ & 49 & 42 & 39 & 10 & 0.4/1 & 0.6/3 & 30.7/75 & 9.6/18 & 10.3/32 & 72.8/159 \\
 & $\mathrm{H_6}$ & 185 & 119 & 110 & 32 & 0.9/1 & 1.2/3 & 64.8/134 & 15.8/18 & 14.3/31 & 129.6/259 \\
 & $\mathrm{LiH}$ & 142 & 108 & 104 & 40 & 0.6/1 & 0.8/3 & 47.5/115 & 12.6/18 & 11.7/31 & 106.2/231 \\
 & $\mathrm{BeH_2}$ & 155 & 122 & 114 & 35 & 0.6/1 & 0.8/3 & 62.5/218 & 11.6/18 & 11.4/31 & 116.9/353 \\
 & $\mathrm{H_2O}$ & 246 & 181 & 176 & 48 & 0.7/1 & 1.0/4 & 85.9/205 & 13.9/18 & 12.6/25 & 151.6/323 \\
\cline{1-12}
\bottomrule
\end{tabular}}
\end{table}

\subsection{NISQ Device Validation}
We validate our numerical expectations against the real-world performance of GALIC compared against QWC, HEC, and FC on a current-generation device: the sparsely connected IBM Kyoto
system. Both the Sherbrooke and Kyoto systems are based on the Eagle r3 QPU, and therefore we expect system behavior to track numerical expectations. We selected the planar $H_4$ molecular Hamiltonian as our problem of interest, as it drew a particular contrast between FC biases and lower-entanglement schemes (HEC, QWC, and GALIC) in spite of its small system size.

For all measurement schemes, sorted insertion grouping resulted in the highest weight group consisting entirely of sparse $Z$ operators, requiring no measurement unitary. Accordingly, we add the known expectation value in post-processing, given its irrelevance to measurement unitary comparison. 

We compare the measurement schemes on a simple Hartree-Fock state to illustrate an initial proof-of-concept, leaving more extensive testing to future work. By using a simple, entanglement-free trial state we eliminate two-qubit errors and minimize thermal noise resulting from state preparation, enabling isolation of the measurement unitaries.

From each selected dataset, we perform randomized sub-sampling with replacement to determine the dependence of bias/variance on the total shot budget. Shots are optimally allocated within the total budget $n$ using:
\begin{equation}
    n_i= n\frac{\sqrt{\mathrm{Var}[O_j]}}{\sum_{j=1}^L\sqrt{\mathrm{Var}[O_j]}},
\end{equation}
where the variance of each commuting group $O_j$ was classically computed. See~\ref{appdx:shot_lagrange} for the derivation using Lagrange multipliers. Adaptive schemes exist to estimate operator variances for classically intractable systems, see Refs~\cite{wu_overlapped_2023,yen_deterministic_2023} for examples. 

\begin{figure}[h]
    \centering
    \includegraphics[width=0.55\linewidth]{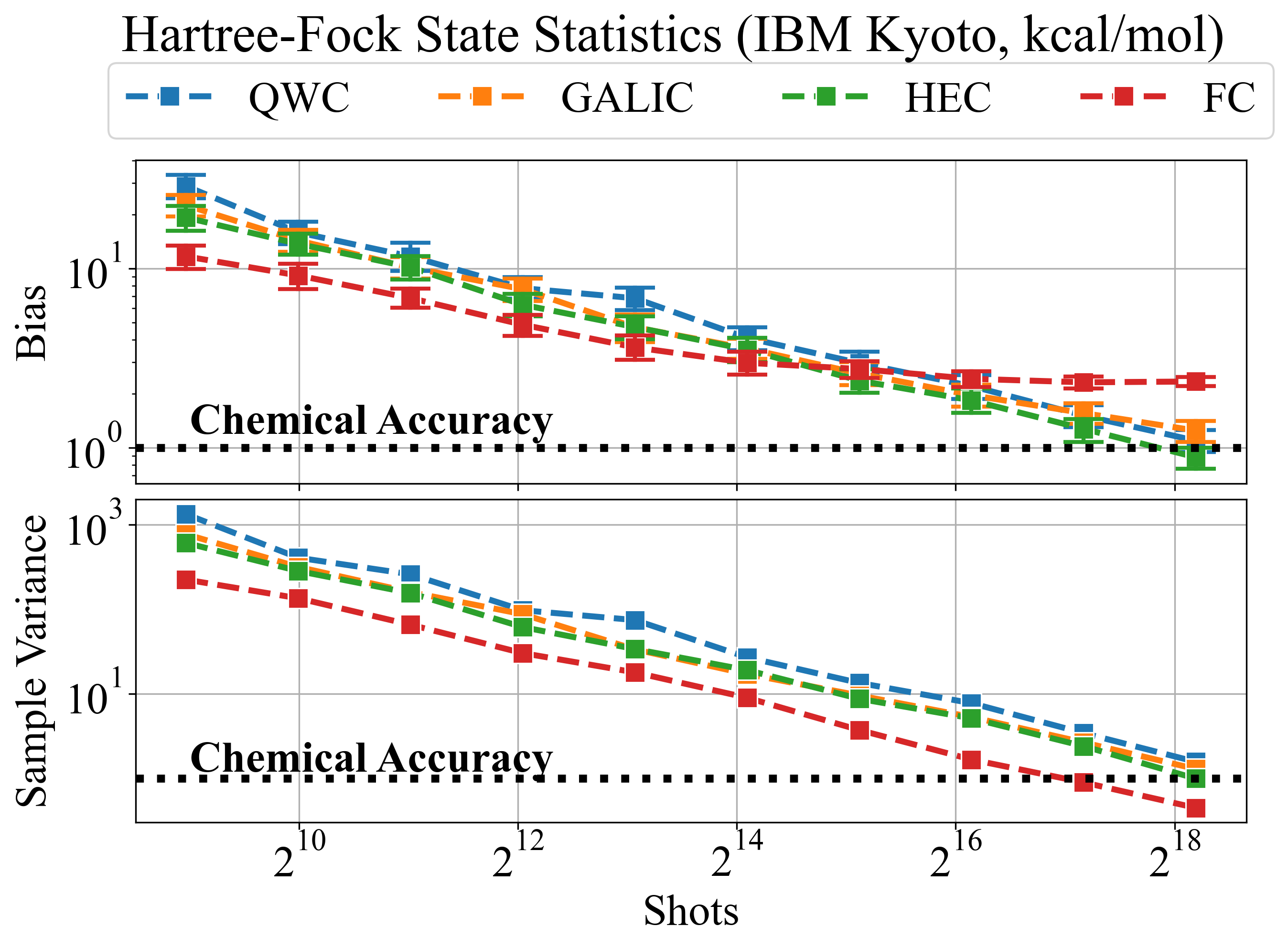}
    \caption{Hartree-Fock state sampling from IBM Kyoto as a function of total shot budget. [Top] Estimator bias (kcal/mol) compared with the true HF energy. [Bottom] Estimator variance (kcal$^2$/mol$^2$). Error bars on biases are computed using 100 sub-samples from a larger 600,000 shot budget dataset.}
    \label{fig:ibm_kyoto_fig}
\end{figure}

Fig.~\ref{fig:ibm_kyoto_fig}[top] shows the energy bias and associated error bars as a function of the total shot budget. While the FC grouping provides advantages for highly shot-limited scenarios, it ultimately plateaus at 2 kcal/mol, higher than the other grouping strategies. The other schemes (HEC, GALIC, and QWC) were within margin-of-error, with an approximate mean of $1.15$ kcal/mol ($1.8$ mHartree), agreeing well with our numerical testing in Table~\ref{tab:bias_var}.

The advantage of grouping schemes lies in lower estimator variances, shown in Fig.~\ref{fig:ibm_kyoto_fig}[bottom]. The FC grouping has a clear advantage in sample-efficient estimation, with an average 2.1$\times$ lower sampling error compared to QWC. However, both GALIC ($1.2\times$) and HEC ($1.27\times$) demonstrated the advantages over QWC expected from numerical testing.

\section{Discussion}\label{sec:discuss}

In this section, we explore the impact of device connectivity and gate fidelity on measurement protocols in detail. Specifically, we observe that near-term quantum devices face a tradeoff between noise and connectivity. For example, in superconducting architectures, increasing the number of neighboring qubits can lead to frequency collisions, crowding, and increased crosstalk~\cite{gao_practical_2021}. Likewise, in trapped-ion systems, although qubits are fully connected, increasing the number of qubits in a single trap can extend the duration of entangling gates, making them more susceptible to thermal noise~\cite{kaushal2020shuttling}.

As demonstrated in previous sections, GALIC effectively leverages both noise and topology information to inform variance-lowering grouping strategies, providing a pathway to navigate the design space between QWC and FC groupings. For this reason, we use GALIC as a tool for our investigation. We note that fidelity and device connectivity also influence algorithm performance factors such as circuit depth and parallelism~\cite{holmes_impact_2020, leymann_bitter_2020}. Although our focus is on estimator bias and sample variance in the measurement scheme sorely, our findings may offer valuable insights for the design of future quantum architectures and algorithms.

\subsection{Estimator Variance}
We begin by evaluating the impact of device design on GALIC variances with an ideal trial state preparation $\rho=\dyad{\psi}$.

\begin{figure}[h]
    \centering
    \includegraphics[width=0.5\linewidth]{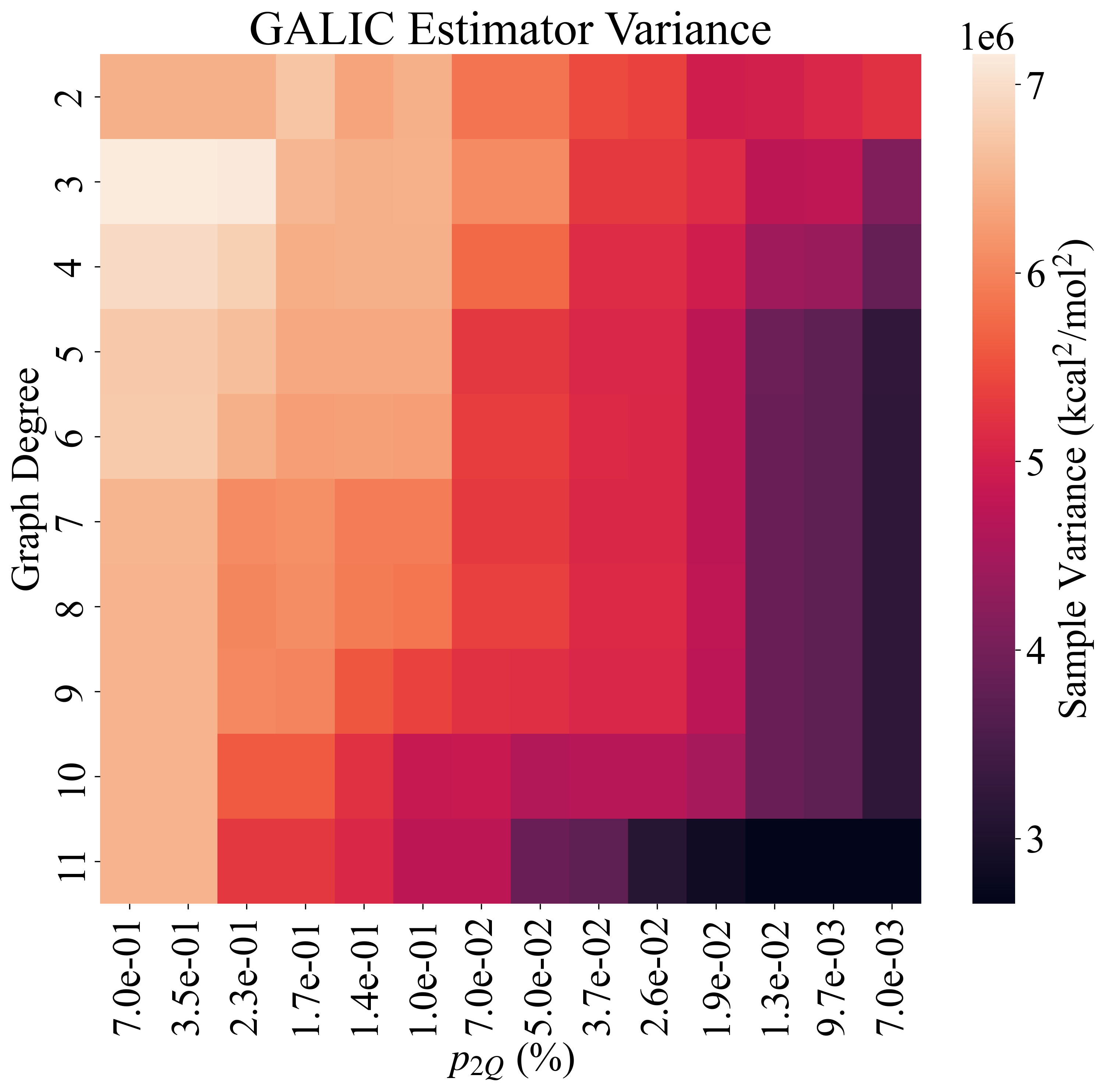}
    \caption{GALIC sample variance (kcal$^2$/mol$^2$) for the $\mathrm{H_6}$ Hamiltonian as a function of device coupling degree (y-axis) and two qubit gate error (\%) (x-axis).}
    \label{fig:h6_heatmap}
\end{figure}

To empirically determine the tradeoff within the GALIC framework, we generate a sequence of coupling graphs with increasing degree, beginning from a ring (degree 2) to a fully connected device (degree 11) and a sequence of device models with increasing fidelity. For each topology/pair we generate a random-regular graph with the desired connectivity. Targeting 1\% relative error, Fig.~\ref{fig:h6_heatmap} shows the average sample variance $\varepsilon^2$ over 10 random quantum states. We see significant reductions in variance from increasing connectivity as well as from decreasing error, however, the latter has a more significant impact.

We quantify the effect of each factor by running a linear regression over the following two models:
\begin{align}
\varepsilon^2 & =\alpha_d d_G+\beta_d, \nonumber \\
\varepsilon^2 & =\alpha_r \log_{10}{r}+\beta_r, \nonumber 
\end{align}
for each row/column of the simulated results, where $d_G$ is the degree of the graph and $r=0.7/p_{\text{2Q}}$ is the noise level reduction. We find $\alpha_d=-0.36\pm 0.003$, $\alpha_r=-3.82\pm 0.01$ for the linear slopes of $d_G$ and $\log{r}$ respectively. The apparent strength of the correlation between noise and  GALIC estimator variance is then about 10.6$\times$ greater than connectivity. Moreover, evaluating the average Pearson correlation coefficient for each yields $p_{d}=-0.73$ and  $p_{r}=-0.93$, indicating a more robust linear relationship as well as a steeper fit. Our numerical testing suggests that device fidelity is a much more significant factor in GALIC shot reductions compared to connectivity.

\subsection{Holistic Noise Analysis}
Estimator variance is not the only reason to consider fidelity over connectivity, as we now show. In previous sections, we assume that the trial state has been prepared perfectly intact, and that circuit measurements constitute the only errors. While useful for comparing measurement schemes, this is far from a realistic assumption. In this sub-section, we reuse the device design space shown in  Fig.~\ref{fig:h6_heatmap}, however, this time we compare the estimator bias when sampling from a shallow ansatz. Each coupling graph was generated as a random regular graph using NetworkX~\cite{hagberg2008exploring}, with circuits transpiled using Qiskit~\cite{qiskit2024}. Both ansätze and measurement unitaries are simulated using density matrix simulations modeling thermal and depolarizing noise. Rather than lowering the two-qubit gate error alone, we adjust all noise parameters in tandem. The error probabilities are scaled down by $r$:
\begin{align}
        p_{2Q,r}&=\frac{p_{2Q,1}}{r},\quad p_{1Q,r}=\frac{p_{1Q,1}}{r},\\
\end{align}
while the thermal coherence times are increased by $\log r$,
\begin{align}
        T_{1/2,r}&=\begin{cases}
            T_{1/2,1}& r=1\\
            T_{1/2,1}\cdot \log r& r>1\\
        \end{cases}\\
\end{align}
where the $r=1$ parameters are the IBM Sherbrooke baseline values (Table~\ref{tab:noise_parameters}).

\begin{figure}[h!]
\begin{subfigure}{0.5\textwidth}
    \centering
    \includegraphics[width=\linewidth]{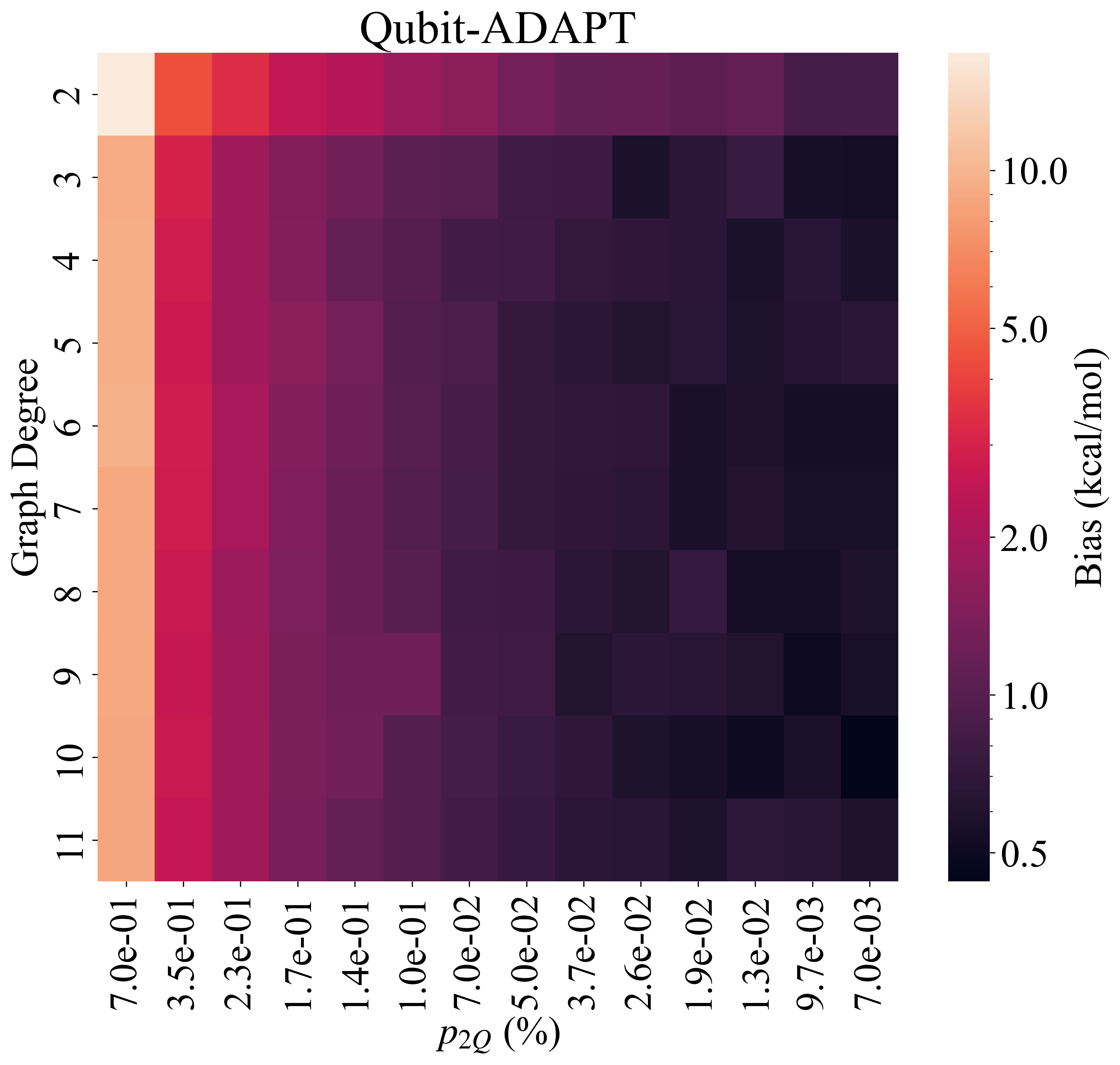}
    \subcaption{}
    \label{fig:enter-label}
\end{subfigure}
\begin{subfigure}{0.5\textwidth}
    \centering
    \includegraphics[width=\linewidth]{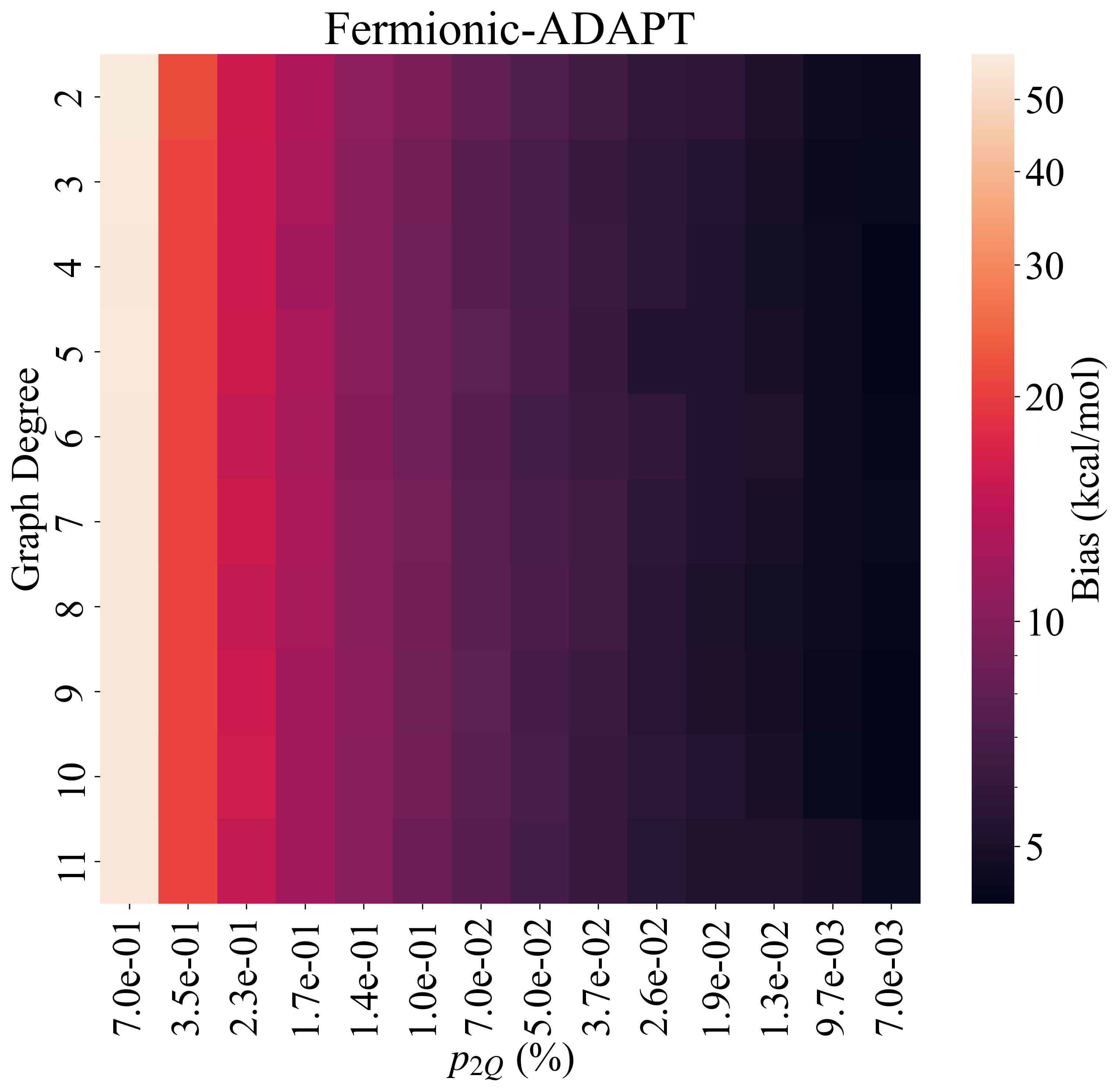}
    \subcaption{}
    \label{fig:fermionic_ansatz}
\end{subfigure}
\caption{Impact of device connectivity and fidelity on GALIC $\mathrm{H_6}$ estimator bias. In (a), we plot 3-iteration Qubit-ADAPT VQE Ansatz estimator bias (Unrouted depth 43, CNOT count 18). In (b), we show 3-iteration Fermionic-ADAPT VQE Ansatz estimator bias (Unrouted depth 668, CNOT count 208).}
\label{fig:whole-circuit-noise}
\end{figure}

Fig.~\ref{fig:whole-circuit-noise} shows the bias dependence on device connectivity and fidelity for two shallow ans\"{a}tze constructed using 3 iterations of adaptive VQE: (a) Qubit-ADAPT-VQE~\cite{tang_qubit-adapt-vqe_2021} algorithm and (b) Fermionic-ADAPT~\cite{grimsley_adaptive_2019}. We report the bias relative to the ideal expectation value of each ansatz, as neither was sufficiently deep to reach the Hamiltonian ground state.

As in the GALIC sample variance testing, device fidelity is a more significant factor than connectivity. However, the bias scaling in Fig.~\ref{fig:whole-circuit-noise} is almost completely dominated by the device noise model, with little dependence on the device topology past $d=2$.

We attribute the lack of dependence on topology in part to transpiler optimizations in layout and routing. The ring ($d=2$) topology incurred $\sim$10\% CNOT overhead from SWAP routing, however $d\geq 3$ required no extra CNOT gates to implement either ansatz. We note that this is a limitation of our testing methodology, as the 6-orbital ansätze were amenable to routing on lower connectivity graphs. Transpilation testing with 15 orbital systems showed $>$10\% increases in 2-qubit gate counts up to degree 6 coupling graphs. However, we conjecture that the device fidelity will remain the dominant effect even in larger systems. Given the engineering challenges involved in manufacturing high-connectivity transmon systems~\cite{gao_practical_2021}, increasing qubit fidelity 
seems a more realistic (though still daunting) path. 

Higher connectivity devices may offer other benefits, such as native support for QEC codes. Variants of the surface code ``only'' require degree $\leq 4$~\cite{gao_practical_2021,fowler_surface_2012}. As our findings show diminishing bias/variance benefit from scaling devices beyond $d=4$, low-connectivity/high-fidelity devices seem a more fruitful direction for quantum chemistry applications.

Biases in Fig.~\ref{fig:fermionic_ansatz} remained above chemical accuracy despite 2 order-of-magnitude reductions in device noise. GALIC prevents excessive measurement bias, but it does not mitigate fidelity loss in the trial state. Error mitigation and (eventually) error correction will be necessary to remedy trial state preparation errors and provide chemically accurate results. QEM schemes incur measurement overhead for more complex estimators and error cancellation techniques, typically trading a decrease in bias for an increase in variance~\cite{cai_quantum_2023}. As shown in our numerical tests, GALIC offers a favorable bias/variance tradeoff which would synergize well with QEM schemes on near-term devices.
\section{Conclusion}

In this work, we introduce a general framework for interpolating between QWC and FC simultaneous measurement along with a novel grouping strategy, GALIC, targeting near-term quantum systems. We show how our framework naturally incorporates previous hardware-efficient proposals~\cite{miller2022hardware,escudero_hardware-efficient_2023} while allowing for easy augmentation with device noise models.

Simulated results demonstrate that GALIC is capable of maintaining chemical accuracy across NISQ architectures while lowering estimator variance compared to qubit-wise commuting groups. Moreover, our approach obtains a more favorable bias-variance tradeoff compared to existing hardware efficient schemes. We validate our results on current NISQ devices, demonstrating minimal bias over QWC on IBM with $1.2\times$ lower sampling error. Our framework also offers insights into design directions for future NISQ devices, as discussed in Sec.~\ref{sec:discuss}.

While low-bias measurement circuits are a necessary condition for maintaining chemical accuracy, they are not sufficient. Crucially, they cannot undo errors in the occurring in the underlying unitary, as we demonstrate in Sec.~\ref{sec:discuss}. However, our technique has the potential to synergize well with existing QEM techniques~\cite{cai_quantum_2023}, such as subspace expansion~\cite{mcclean_hybrid_2017,mcclean_decoding_2020} and zero noise extrapolation~\cite{he_zero-noise_2020,temme2017error}, which correct biases at the cost of more measurements.

There are numerous directions for future work to build on our proposals. The proposed GALIC scheme uses static partitioning without covariance information. Numerous works have shown benefits from adaptive covariance estimation and shot allocation~\cite{yen_deterministic_2023, wu_overlapped_2023}. GALIC also considers a single parameter model of device fidelity (mean/median depolarizing error), while many providers release per-coupler calibration data. A finer-grained GALIC approach may lead to further optimized grouping schemes with high-fidelity estimates of noise parameters~\cite{zheng_bayesian_2023}.

Our grouping function framework also expands the design space of simultaneous measurement schemes. Potential research directions include problem-informed context functions, grouping functions with real-valued codomains for grouping optimization, and adaptive parameterizations. Our proposed framework may also be useful in formal analyses of grouping heuristics, as our partial ordering can provide lower/upper bounds on optimality across different grouping functions.

\section*{Acknowledgement}

We thank Nathan Wiebe and Muqing Zheng for helpful conversations during data collection and analysis. This work was supported by the ``Embedding QC into Many-body Frameworks for Strongly Correlated Molecular and Materials Systems'' project, which is funded by the U.S. Department of Energy, Office of Science, Office of Basic Energy Sciences (BES), the Division of Chemical Sciences, Geosciences, and Biosciences (under award 72689). KK and AL also acknowledge the support from the U.S. Department of Energy, Office of Science, National Quantum Information Science Research Centers, Quantum Science Center (QSC). This research used resources of the Oak Ridge Leadership Computing Facility (OLCF), which is a DOE Office of Science User Facility supported under Contract DE-AC05-00OR22725. This research used resources of the National Energy Research Scientific Computing Center (NERSC), a U.S. Department of Energy Office of Science User Facility located at Lawrence Berkeley National Laboratory, operated under Contract No. DE-AC02-05CH11231. The Pacific Northwest National Laboratory (PNNL) is operated by Battelle for the U.S. Department of Energy under Contract DE-AC05-76RL01830.
\appendix
\section{Sorted Insertion}\label{appdx:si}
In this section we overview the Sorted Insertion (SI) heuristic used for constructing commuting groups. SI was first introduced in Ref.~\cite{crawford_efficient_2021} as a computationally efficient and high-performance alternative to clique cover algorithms used in previous works~\cite{gokhale_optimization_2020,gokhale2020n}. Our generalized SI, which takes both the weighted operator list $O$ and a generalized grouping function $f$ with context variables $\theta_1, \theta_2,...$, is shown in Algorithm~\ref{alg:sorted_insertion}. Simply put, Sorted Insertion greedily constructs a set of cliques, where operators with higher-magnitude coefficients are more likely to belong to a larger commuting group.  

\begin{algorithm}
    \begin{algorithmic}[1]
        \Procedure{Sorted Insertion}{Operators $O=(c_1P_1,c_2P_2,...,c_MP_M)$, Grouping Function $f(\cdot,\theta_1,\theta_2...)$}
        \State Sort $O$ in descending weight order
        \State Set $Groups\leftarrow\emptyset$\Comment{Set of commuting groups}
        \State Set $Added\leftarrow\emptyset$\Comment{Set of Paulis added to groups}
        \While{$|Added|< M$}\Comment{While some Paulis not added to a commuting set}
        \State $G=\emptyset$
        \For{$P_i$ in $O \setminus A$}\Comment{Iterate over Paulis in sorted order}
            \If{$f(G\cup \{P_i\}, \theta_1,\theta_2...) = 1$}\Comment{Grouping function check}
            \State $G\leftarrow G\cup\{P_i\}$\Comment{Add $P_i$ to the current group}
            \State $Added\leftarrow Added\cup\{P_i\}$ \Comment{Remove $P_i$ from consideration for future groups}
            \EndIf
        \EndFor
        \State $Groups\leftarrow Groups\cup\{G\}$\Comment{Add $G$ to the set of commuting groups}
        \EndWhile
        \State \textbf{return} $Groups$
        \EndProcedure
    \end{algorithmic}
    \caption{Sorted Insertion for Generalized, Context-Aware Grouping Functions}
    \label{alg:sorted_insertion}
\end{algorithm}

Ref.~\cite{yen_deterministic_2023} augmented SI to allow for overlapping groups, however it has been shown that simultaneous measurement using overlapping groups are particularly sensitive to covariances between terms~\cite{wu_overlapped_2023,yen_deterministic_2023}, requiring iterative covariance estimates. Our proposal is entirely compatible with preexisting adaptive algorithms, and we leave further comparison to future work.
\section{Shot Allocation by Lagrange Multipliers}\label{appdx:shot_lagrange}
\newcommand{\Lag}{\mathcal{L}}
Lagrange multipliers are commonly used in shot allocation schemes to minimize shot overhead within precision tolerances~\cite{zhu_optimizing_2024,crawford_efficient_2021,gresch_guaranteed_2023}. There are common two formulations leading to the same outcome, and we recount both for completeness.

In the first formulation, we wish to minimize estimator variance $\sum_{i=1}^{L}\frac{\Var{O_i}}{n_i}$ subject to the constraint $\sum_{i=1}^L n_i=n$ for fixed shot budget $n$~\cite{zhu_optimizing_2024}. We then have the formal problem:
\begin{equation}
    \begin{split}
        &\min \sum_{i=1}^{L}\frac{\Var{O_i}}{n_i}\leq \varepsilon^2\\
        &\mathrm{s.t.} \sum_{i=1}^Ln_i\leq n
    \end{split}
\end{equation}
with the associated Lagrangian $\Lag$:
\begin{equation}
    \Lag(\vec n, \lambda)=\sum_{i=1}^{L}\frac{\Var{O_i}}{n_i} +\lambda\cdot ( \sum_{i=1}^{L}n_i-n)
\end{equation}
where $\lambda\in\mathbb{R}$ is a Lagrange multiplier. We can obtain the minimizer of $\Lag$ by applying the KKT conditions:
\begin{align}
\frac{\partial\L}{\partial n_i}=0 \nonumber \\
\sum_{i=1}^{L}n_i-n = 0 \nonumber
\end{align}
From the first KKT condition we obtain:
\begin{equation}
    \begin{split}
        -\frac{\Var{O_i}}{n_i^2}+\lambda&=0\\
        \frac{1}{n_i}&=\frac{\sqrt{\lambda}}{\sqrt {\Var O_i}}\\
        n_i&=\frac{\sqrt {\Var O_i}}{\sqrt{\lambda}}\\
    \end{split}
\end{equation}
We then substitute $n_i$ into the second KKT condition to obtain $\lambda$:
\begin{align}
 \sum_{i=1}^{L}n_i-n & = 0 \nonumber \\
 \sum_{i=1}^{L}\frac{\sqrt {\Var O_i}}{\sqrt{\lambda}}-n & = 0 \nonumber \\
\frac{1}{\sqrt{\lambda}}-n & = \frac{n}{\sum_{i=1}^{L}\sqrt {\Var O_i}} \nonumber
\end{align}
yielding the minimal-variance shot assignment:
\begin{equation}
    n_i=n\frac{\sqrt {\Var O_i}}{\sum_{j=1}^{L}\sqrt {\Var O_j}}
\end{equation}

In the second formulation, we wish to minimize the number of shots $n$ subject to a minimum precision $\varepsilon$~\cite{crawford_efficient_2021}. The optimization problem can be formally stated as:
\begin{equation}
    \begin{split}
        &\min n=\sum_{i=1}^Ln_i\\
        &\mathrm{s.t.} \sum_{i=1}^{L}\frac{\Var{O_i}}{n_i}\leq \varepsilon^2
    \end{split}
\end{equation}
Adding a Lagrange multiplier $\lambda$ to the constraint gives the Lagrangian $\Lag$:

\begin{equation}
    \Lag(\vec n, \lambda)=\sum_{i=1}^Ln_i+\lambda\cdot ( \sum_{i=1}^{L}\frac{\Var{O_i}}{n_i}-\varepsilon^2)
\end{equation}
Applying the KKT conditions yields:
\begin{align}
\frac{\partial\L}{\partial n_i}=0 \nonumber \\
 \sum_{i=1}^{L}\frac{\Var{O_i}}{n_i}-\varepsilon^2= 0 \nonumber \\
\end{align}
From the first condition, we obtain:
\begin{equation}\label{eqn:n_i_lambda}
\begin{split}
    1 -\lambda\frac{\Var{O_i}}{n_i^{2}}&=0\\
    \frac{1}{n_i}&=\frac{1}{\sqrt{\lambda}\sqrt{\Var{O_i}}}
\end{split}    
\end{equation}

Plugging each $1/n_i$ into the second KKT condition yields:
\begin{equation}
\begin{split}
    \sum_{i=1}^{L}\frac{\Var{O_i}}{\sqrt{\lambda}\sqrt{\Var{O_i}}}-\varepsilon^2&= 0\\
    \sum_{i=1}^{L}\frac{\sqrt{\Var{O_i}}}{\sqrt{\lambda}}&=\varepsilon^2\\
    \frac{1}{\sqrt{\lambda}}&=\frac{\varepsilon^2}{\sum_{i=1}^{L}\sqrt{\Var{O_i}}}
\end{split}
\end{equation}
Finally, substituting back into Eqn.~\ref{eqn:n_i_lambda}:
\begin{equation}
\begin{split}
    \frac{1}{n_i}&=\frac{\varepsilon^2}{\sqrt{\Var{O_i}}\sum_{j=1}^{L}\sqrt{\Var{O_j}}}\\
    n_i&=\frac{1}{\varepsilon^2}\sqrt{\Var{O_i}}\sum_{j=1}^{L}\sqrt{\Var{O_j}}
\end{split}
\end{equation}
or, expressing the total shot budget $n=\sum_{i=1}^Ln_i=\frac{1}{\varepsilon^2}\left(\sum_{i=1}^{L}\sqrt{\Var{O_i}}\right)^2$, we obtain the minimum shot count needed to achieve $\varepsilon$ precision:
\begin{equation}
    n_i=n\frac{\sqrt{\Var{O_i}}}{\sum_{j=1}^{L}\sqrt{\Var{O_j}}}
\end{equation}
Both lead to equivalent outcomes, and depend solely on whether the use case calls for a restriction on the total number of shots (first case) or a restriction on total estimator accuracy (second case).


\section*{References}
\bibliographystyle{unsrt}
\bibliography{refs}

\end{document}